\documentclass[journal]{IEEEtran}
\usepackage{epsfig,amsmath,amssymb,epsf,amsthm,scalefnt,multirow}
\usepackage{xcolor}
\usepackage{float}
\usepackage[caption=false,font=footnotesize,textfont=sf]{subfig}
\usepackage{textcomp}
\usepackage{stfloats}
\usepackage{cite}
\usepackage{psfrag}
\usepackage{bm}
\usepackage{algorithmic,algorithm}
\usepackage{graphicx}
\usepackage{mleftright}

\newtheorem{theorem}{Theorem}
\newtheorem{lemma}{Lemma}
\newtheorem*{lemma*}{Lemma}

\newtheorem{corollary}{Corollary}

\newcommand{\norm}[1]{\left\lVert#1\right\rVert}

\makeatletter
\newcommand{\vast}{\bBigg@{2.8}}
\newcommand{\Vast}{\bBigg@{4}}
\makeatother

 \allowdisplaybreaks
\begin{document}

\title{Spectrum Sharing Between Low Earth Orbit Satellite and Terrestrial Networks:\\ A Stochastic Geometry Perspective Analysis}

%\author{Daeun~Kim,~\IEEEmembership{Student~Member,~IEEE}, Jeonghun~Park~\IEEEmembership{Senior~Member,~IEEE} ~and~Namyoon~Lee,~\IEEEmembership{Senior~Member,~IEEE}

\author{Daeun~Kim, Jeonghun~Park, Jinseok~Choi, and Namyoon~Lee
% \author{Jeonghun~Park, Jinseok~Choi, and Namyoon~Lee

\thanks{D. Kim is with the Department of Electrical Engineering, POSTECH, Pohang, Gyeongbuk 37673, South Korea (email: {\texttt{daeun.kim@postech.ac.kr)}}). J. Park is with School of Electrical and Electronic Engineering, Yonsei University, South Korea (e-mail:{\texttt{jhpark@yonsei.ac.kr}}). J. Choi is with School of Electrical Engineering, KAIST, South Korea (email:{\texttt{jinseok@kaist.ac.kr}}), N. Lee is with Department of Electrical Engineering, Korea University, South Korea (e-mail:{\texttt{namyoon@korea.ac.kr}}). 
}
}

% \thanks{D. Kim is with the Department of Electrical Engineering, POSTECH, Pohang, Gyeongbuk 37673, South Korea  (e-mail: daeun.kim@postech.ac.kr).}
% 	\thanks{N. Lee is with the School of Electrical Engineering, Korea University, Seoul 02841, South Korea (e-mail: namyoon@korea.ac.kr).}
% }
\maketitle

\begin{abstract} 
Low Earth orbit (LEO) satellite networks with mega constellations have the potential to provide 5G and beyond services ubiquitously. However, these networks may introduce mutual interference to both satellite and terrestrial networks, particularly when sharing spectrum resources. In this paper, we present a system-level performance analysis to address these interference issues using the tool of stochastic geometry. We model the spatial distributions of satellites, satellite users, terrestrial base stations (BSs), and terrestrial users using independent Poisson point processes on the surfaces of concentric spheres. Under these spatial models, we derive analytical expressions for the ergodic spectral efficiency of uplink (UL) and downlink (DL) satellite networks when they share spectrum with both UL and DL terrestrial networks. These derived ergodic expressions capture comprehensive network parameters, including the densities of satellite and terrestrial networks, the path-loss exponent, and fading. From our analysis, we determine the conditions under which spectrum sharing with UL terrestrial networks is advantageous for both UL and DL satellite networks. Our key finding is that the optimal spectrum sharing configuration among the four possible configurations depends on the density ratio between terrestrial BSs and users, providing a design guideline for spectrum management. Simulation results confirm the accuracy of our derived expressions.\end{abstract}

\section{Introduction}
% With their capacity for broad coverage and connectivity, satellite networks are promising solutions for providing global coverage . Recently,

Low Earth orbit (LEO) satellite networks with mega constellations have garnered significant interest from both industry and academia due to their ability to provide lower latency and higher data rates compared to traditional GEO satellite networks \cite{Kodheli2021, Zhu2022, Homssi2022}. Additionally, LEO satellites offer a complementary solution for extending the coverage of terrestrial networks, particularly in remote areas where terrestrial networks are sparse or non-existent.

Despite the benefits LEO satellite networks bring to global connectivity, they also introduce previously unseen mutual interference problems, particularly when sharing spectrum with terrestrial networks. The growing demand for broadband communication services poses significant challenges to the limited and congested spectrum resources. The trend of using higher frequency bands, such as the Ka-band, in cellular networks indicates that these bands are shared with satellite networks, and numerous terrestrial terminals potentially generate significant aggregated interference to satellite networks \cite{Lim2024}. Furthermore, Starlink plans to utilize the S-band for direct-to-cellular service, leading to the coexistence of current cellular and satellite networks within the same band, which is already heavily occupied \cite{FCC_Sband}.

% However, despite the benefits LEO satellite networks bring to global connectivity, they also introduce new challenges, particularly concerning spectrum management. The growing demand for broadband communication services poses significant challenges to the limited and congested spectrum resources. Spectrum sharing between satellite and terrestrial networks presents an appealing solution to this problem. By allowing different systems to operate on the same frequency bands, spectrum sharing can alleviate spectrum scarcity and improve spectrum efficiency.  

Spectrum sharing introduces inter-system interference that can severely impact both networks. As a result, identifying efficient spectrum-sharing configurations is crucial in the design and operation of satellite networks. Specifically, there are four possible spectrum-sharing configurations between satellite and terrestrial networks:
\begin{itemize}
    \item Configuration 1: UL satellite and DL terrestrial networks,
 \item Configuration 2: UL satellite and UL terrestrial networks,
 \item Configuration 3: DL satellite and DL terrestrial networks,
  \item Configuration 4: DL satellite and UL terrestrial networks.
\end{itemize}

Although many prior works have extensively studied the coexistence and spectrum sharing problems between satellite and terrestrial networks \cite{Hao2021, Hao2024, Cho2020, Jumaily2022, 3GPP863}, none have definitively determined which configuration is optimal under varying network parameter conditions. A major limitation of prior approaches is that they typically require complex system-level simulations to assess performance across various system parameters and account for different sources of randomness. Consequently, obtaining insights for network planning based on these parameters can be challenging. Therefore, an analytical tool is necessary to characterize the performance of satellite networks and provide guidance for satellite network deployment under spectrum sharing with terrestrial networks.

In this paper, we address this gap by deriving analytical expressions for the UL and DL performance of LEO satellite networks under four possible spectrum-sharing configurations using the tool of stochastic geometry. This approach allows us to determine the conditions under which spectrum sharing with UL terrestrial networks is advantageous for both UL and DL satellite networks. Our analysis provides a comprehensive understanding of the optimal spectrum-sharing configuration for dense LEO satellite and terrestrial networks, facilitating more efficient network planning and deployment.

% In this paper, we address the question: What is the optimal spectrum sharing configuration between dense LEO satellite and terrestrial networks? To answer this, we first derive analytical expressions for UL and DL LEO satellite networks under the four possible spectrum sharing configurations using the tool of stochastic geometry. Through this analysis, we determine the conditions under which spectrum sharing with UL terrestrial networks is advantageous for both UL and DL satellite networks.

\subsection{Related Works}
The coexistence and spectrum sharing between satellite and terrestrial networks have been extensively studied \cite{Hao2021, Hao2024, Cho2020, Jumaily2022, 3GPP863}. In \cite{Hao2021}, co-channel interferences were investigated under various spectrum-sharing mechanisms. Furthermore, the interference-aware radio resource group sharing mechanism was proposed as an interference mitigation technique in \cite{Hao2024}. In \cite{Cho2020}, the accumulated interference power from numerous 5G systems to a satellite receiver was evaluated. An evaluation of 5G interference with fixed satellite services was provided in \cite{Jumaily2022}. In \cite{3GPP863}, simulation results were presented for average throughput loss originating from the coexistence between terrestrial networks and non-terrestrial networks in the S-band. However, these approaches require complex system-level simulations to assess performance across various system parameters and account for different sources of randomness.

%Consequently, it is challenging to obtain insights for network planning based on the network parameters. Therefore, an analytical tool is necessary to characterize the performance of satellite networks and provide guidance for the satellite network deployment under spectrum sharing with terrestrial networks.

Stochastic geometry is an effective tool for characterizing the spatially averaged performance of various wireless networks. By using Poisson point processes (PPPs) to model the locations of base stations (BSs) and users, this approach has yielded valuable insights into the coverage and rate performances of various cellular network scenarios \cite{Guo2013, Andrews2011, Dhillon2012, Heath2013, Di2016, Novlan2013, Lee2015, Lee2015-2, Park2016, Singh2015}. Furthermore, PPP-based performance analysis for LEO satellite networks was introduced in \cite{Okati2022, Park2023, Chae2023, Park2023_2, Lee2022, Kim2024, Yim2024}. Analytical expressions for the DL coverage probability and average data rate of a massive inclined LEO constellation were provided in \cite{Okati2022}, using a nonhomogeneous PPP to account for the inherent non-uniform distribution of satellites across different latitudes. In \cite{Park2023,Lee0}, DL coverage expressions were introduced by modeling the locations of satellites and users using PPP on the surfaces of concentric spheres, demonstrating that optimal satellite density decreases logarithmically with satellite altitude. The coverage probability and ergodic rate were derived as functions of beamwidth, revealing the impact of system parameters on optimal beam control in \cite{Chae2023}. Furthermore, in \cite{Park2023_2}, a unified network model for satellite-terrestrial integrated networks was provided, exploring the benefits of coverage extension and data offloading. The coverage probability expression for LEO satellite networks focusing on orbit geometry parameters was derived in \cite{Lee2022,Choi1,Choi2}. In \cite{Kim2024}, coverage probability expressions considering dynamic coordinated beamforming were provided. Additionally, the DL coverage performance of multi-tier integrated satellite-terrestrial networks was analyzed in \cite{Yim2024}.

Furthermore, the performance analysis of satellite networks using stochastic geometry tools under spectrum sharing or coexistence with terrestrial networks has been investigated in several studies \cite{Lim2024, Kolawole2017, Yastrebova2020, Okati2024}. In \cite{Lim2024}, the cumulative distribution function (CDF) of interference from terrestrial nodes to the satellite was derived by modeling the terrestrial nodes with a PPP, and limits on terrestrial node density to protect satellite services were analyzed. The DL outage performance at the satellite user was provided in \cite{Kolawole2017} by considering interference from terrestrial BSs modeled with a PPP, and the effects of terrestrial transmission schemes on satellite networks were compared. In \cite{Yastrebova2020}, the UL coverage probability of LEO satellite networks was derived by considering terrestrial interference to the satellite in a frequency reuse scenario. Additionally, the coverage and rate expressions of DL terrestrial networks were provided in \cite{Okati2024} for two coexistence scenarios, by modeling satellite users and terrestrial BSs as a binomial point process (BPP). These prior works have analytically investigated the performance of spectrum sharing between satellite and terrestrial networks. Nonetheless, these studies are limited, as they do not address the performance of both UL and DL satellite networks in two spectrum-sharing configurations with terrestrial networks. Additionally, there is a lack of comparison of the spectrum-sharing configurations for specific network parameters and insights into which configurations are more advantageous. This is the main focus of our paper.

%Lim2024:No rate or coverage of satellite and comparison between configurations, just interference analysis but the type of satellite user is crucial for the configuration comparison.
%kolawole2015:not consider the spherical, not consider the multiple satellites, not consider the configuration comparison
%okati2024: just consider one satellite, and no performance for satellite therefore, not spherical.

\subsection{Contributions}
In this paper, we provide analytical expressions for the ergodic spectral efficiency of UL and DL satellite networks, considering spectrum sharing with UL or DL terrestrial networks. The major contributions of this paper are summarized as follows:

\begin{itemize}
    \item  We model the locations of satellites, satellite users, terrestrial BSs, and terrestrial users as independent homogeneous spherical PPPs on the surfaces of concentric spheres. Specifically, multi-layered spheres with varying radii reflect the different altitudes of each node type. We then define the visible areas for various transmitters from the typical satellite for UL satellite networks and from the typical satellite user for DL satellite networks.

    \item To account for interference from terrestrial to satellite networks, we derive the Laplace transform of the aggregated interference power from both terrestrial and satellite networks. To this end, we adopt a Shadowed-Rician fading model for satellite channels and a Nakagami-$m$ fading model for terrestrial channels, which suitably capture the fading effects. For a more realistic investigation of the impact of terrestrial networks on the typical satellite in the satellite UL scenario, we consider a down-tilt BS antenna model, where the BS antenna gain is tuned by considering the elevation angle of the typical satellite from the BSs.
    
    \item Building on the derived Laplace transform of the aggregated interference power, we present exact expressions for the ergodic spectral efficiency for both UL and DL satellite networks under spectrum sharing with terrestrial networks. Our analysis includes deriving the signal-to-interference-plus-noise ratio (SINR) coverage probability, conditioned on the existence of at least one satellite or satellite user for UL and DL networks, respectively. By marginalizing the SINR coverage probability, we obtain the ergodic spectral efficiency expression. The derived expressions maintain full generality in terms of network parameters such as antenna gain, path-loss exponents, transmit power, altitudes, and densities.

    \item Through simulations, we confirm the accuracy of the derived expressions. Additionally, we compare the performance of spectrum sharing with UL and DL terrestrial networks. Specifically, we investigate the ergodic spectral efficiency based on the varying density ratio between terrestrial base stations and terrestrial users. Our key findings indicate that in UL satellite networks, the beneficial spectrum-sharing mode of terrestrial networks depends on the density ratio of terrestrial networks and the type of satellite user. In DL satellite networks, although the advantageous spectrum-sharing mode of terrestrial networks also depends on the density ratio of terrestrial networks, spectrum sharing with DL terrestrial networks shows advantages considering the general node densities of cellular networks.

\end{itemize}

\subsection{Organization}
This paper is organized as follows: Section II describes the satellite network models, including antenna gain and channel fading models, and defines the ergodic spectrum efficiency according to spectrum sharing scenarios between satellite and terrestrial networks. Section III presents key lemmas that serve as building blocks for deriving the analytical expressions for coverage probabilities discussed in the subsequent section. In Section IV, we derive the UL ergodic spectral efficiency of the satellite networks under two different spectrum sharing scenarios. Section V also provides the DL ergodic spectral efficiency for two distinct spectrum sharing scenarios in satellite networks. The paper concludes in Section VI.

\section{System Model}
This section explains the spatial model of satellite networks and the signal models, considering spectrum sharing between satellite and terrestrial networks. We focus on four spectrum-sharing configurations from the satellite network perspective, where the satellite UL and DL networks share spectrum with either the terrestrial DL or UL. Additionally, we establish performance metrics to assess the satellite networks' performance under the four configurations.

\begin{figure}[t]
    \centering
    \subfloat[]{\includegraphics[width=7cm]{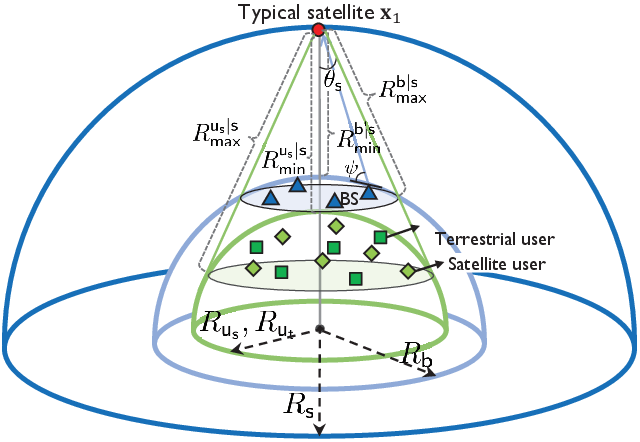}}  
    \\
    \subfloat[]{\includegraphics[width=7cm]{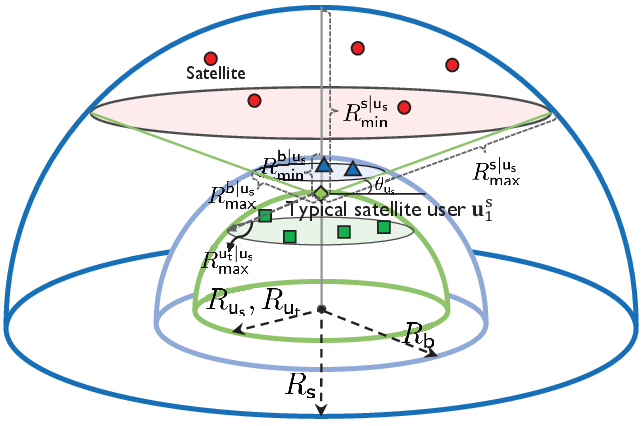}}
    \caption{Illustrations of (a) satellite UL networks and (b) satellite DL networks considering the spectrum sharing with terrestrial networks.} \label{fig:networks}
\end{figure}

\subsection{Satellite and Terrestrial Network Model}
\subsubsection{Spatial Distribution of Satellites, BSs, and Users} 
We assume that the satellites are located on the surface of a sphere with radius $R_{\sf s}$ and satellite users are located on the surface of a concentric sphere with radius $R_{\sf u_s}$. The locations of satellites and satellite users are assumed to follow a homogeneous spherical Poisson point process (SPPP) with densities $\lambda_{\sf s}$ and $\lambda_{\sf u_s}$, denoted as $\Phi_{\sf s}=\{\mathbf{x}_1,\ldots,\mathbf{x}_N\}$ and $\Phi_{\sf u_s}=\{\mathbf{u}^{\sf s}_1,\ldots,\mathbf{u}^{\sf s}_L\}$, respectively. Similarly, we assume that the terrestrial BSs are located on the surface of a concentric sphere with radius $R_{\sf b}$ and terrestrial users are located on the surface of a concentric sphere with radius $R_{\sf u_t}$. The locations of terrestrial BSs and terrestrial users are assumed to be distributed according to the homogenous SPPP with densities $\lambda_{\sf b}$ and $\lambda_{\sf u_t}$, denoted as $\Phi_{\sf b}=\{\mathbf{b}_1,\ldots,\mathbf{b}_M\}$ and $\Phi_{\sf u_t}=\{\mathbf{u}^{\sf t}_1,\ldots,\mathbf{u}^{\sf t}_K\}$, respectively.

\subsubsection{Visible Spherical Cap in UL Satellite Networks} We first consider the typical satellite $\mathbf{x}_1$ located at $(0,0,R_{\sf s})$, by Slivnyak's theorem \cite{Haenggi2005}. Since the BSs and users are located on the surface of spheres, we consider the visible spherical caps which are the visible area of BSs and users to the typical satellite as illustrated in Fig. \ref{fig:networks}-(a). The visible spherical caps $\mathcal{A}_{o|{\sf s}}$ for BSs $(o={\sf b})$ and users $(o={\sf u_s, u_t})$ are determined by a visible angle $\theta_{\sf s}$ from the typical satellite, where BSs and users on the sphere's surface within this angle are visible from the typical satellite. For $o \in \{\sf  u_s, b, u_t\}$, the BSs and users are distributed according to SPPPs with expected numbers of $\lambda_o |\mathcal{A}_{o|\sf s}|$ in the area of visible spherical caps. 

We define the minimum and maximum distances between the typical satellite and visible spherical caps $\mathcal{A}_{o|\sf s}$ as $R_{\sf min}^{o|\sf s}=R_{\sf s}-R_{o}$ and $R_{\sf max}^{o|\sf s}=R_{\sf s}\cos{\theta_{\sf s}}-\sqrt{R_{o}^2-R_{\sf s}^2 \sin^2{\theta_{\sf s}}}$, respectively. Then, the areas of the visible spherical caps are computed by Archimedes' hat theorem \cite{cundy1989} as
\begin{align}
        |\mathcal{A}_{o|\sf s}| = 2\pi R_{o}\left(R_o-\sqrt{R_o^2-(R_{\sf max}^{o|\sf s})^2\sin^2{\theta_{\sf s}}}\right).
\end{align}
 Further, $\mathcal{A}_{o|\sf s}(r)$ represents the spherical cap that includes all points whose distance from the typical satellite is less than $r$.

% When $h_{\sf E}=\frac{R_{\sf E}^2}{R_{\sf s}}$, the $R_{\sf max}=\sqrt{R_{\sf s}^2-R_{\sf E}^2}$ is the distance between the typical satellite and point of tangency. Further, we define the spherical cap $\mathcal{A}_r$, which contains all points whose distance from the typical satellite's location is less than $r$ for $R_{\sf min} \le r \le R_{\sf max}$. Then, the area of the spherical cap $|\mathcal{A}_r|$ is computed by
% \begin{align}
%     |\mathcal{A}_r|=2\pi R_{\sf E}(R_{\sf E}-h_r),
% \end{align}
% where $h_r = \frac{R_{\sf s}^2+R_{\sf E}^2-r^2}{2R_{\sf s}}$ is the distance between the center of the sphere and the plane of spherical cap $\mathcal{A}_r$.

\subsubsection{Visible Spherical Cap in DL Satellite Networks}
We consider the typical satellite user $\mathbf{u}_1^{\sf s}$ at $(0,0,R_{\sf u_s})$. Then, the visible areas from the typical satellite user are determined by an elevation angle $\theta_{\sf u_s}$ of the typical satellite user as depicted in Fig. \ref{fig:networks}-(b). The satellites and BSs are located according to the SPPP within the visible spherical caps $\mathcal{A}_{o|\sf u_s}$ for satellites $(o=\sf s)$ and BSs $(o=\sf b)$, with the expected number of $\lambda_o |\mathcal{A}_{o|\sf u_s}|$. 

The minimum and the maximum distance between the typical satellite user and the visible spherical caps $\mathcal{A}_{o|\sf u_s}$ are computed by $R_{\sf min}^{o|\sf u_s}=R_{o}-R_{\sf u_s}$ and $R_{\sf max}^{o|\sf u_s}=\sqrt{R_o^2-R_{\sf u_s}^2+R_{\sf u_s}^2 \sin^2{\theta_{\sf u_s}}}-R_{\sf u_s}\sin{\theta_{\sf u_s}}$ for $o\in\{\sf s, b\}$, respectively. Then, the areas of the visible spherical caps for satellites and BSs are computed by 
\begin{align}
    |\mathcal{A}_{o|\sf u_s}| = 2\pi R_{o} \left(R_{o}-R_{\sf u_s}-R_{\sf max}^{o|\sf u_s}\sin{\theta_{\sf u_s}}\right).
\end{align}
 Similarly, $\mathcal{A}_{o|\sf u_s}(r)$ represents the spherical cap that includes all points whose distance from the typical satellite user is less than $r$.

On the other hand, the terrestrial user is not visible from the typical satellite user since their altitude is defined as the same. However, in reality, the nearby users are interfered with each other. To reflect this, we define the visible area of terrestrial users, denoted as $\mathcal{A}_{\sf u_t|u_s}$, as a circular area centered around the typical satellite user with a radius of $R_{\sf max}^{\sf u_t|u_s}$. Then, the area of the visible region for the terrestrial user is given by
\begin{align}
     |\mathcal{A}_{\sf u_t| u_s}| = \pi \left(R_{\sf max}^{\sf u_t|u_s}\right)^2.
\end{align}

\subsection{Beamforming Gain Model}
To characterize the beamforming gain of satellites, satellite users, BSs, and terrestrial users, we adopt the sectored antenna gain model, where the antenna gain is simplified into a rectangular pattern \cite{Dabiri2021}. Using sectored antenna gain approximation, we model the antenna gain of satellites with two lobes of main lobes $G_{\sf s}^{\sf main}$ and side lobes $G_{\sf s}^{\sf side}$. Similarly, we employ the two-lobe approximation to model the antenna gain of both satellite users ($o={\sf u_s}$) and terrestrial users ($o={\sf u_t}$), with $G_{o}^{\sf main}$ and $G_{o}^{\sf side}$ denoting the main and side lobes of antenna gain, respectively. 

On the other hand, the antenna gain of BS to the satellite is highly affected by the satellite's elevation angle due to the slight downward tilt of BS antennas. Therefore, the two-lobe approximation for the beam gain of the BS antenna significantly simplifies the side-lobe effects relative to the elevation angle. For example, for the elevation angle above $40^{\circ}$, the side-lobe beam gain to the satellite significantly diminishes \cite{kang2023}. To address this, we model the beam gain of BS to the satellite with three levels by dividing the side-lobe beam gain $G_{\sf b}^{\sf side}$ into two levels of $G_{\sf b_H}^{\sf side}$ and $G_{\sf b_L}^{\sf side}$ as follows:
\begin{align}
        G_{\sf b} =\begin{cases}
   G_{\sf b}^{\sf main}  & 0 \le \psi \le \psi_1^{\sf th} \\
   G_{\sf b_H}^{\sf side}  & \psi_1^{\sf th} < \psi \le \psi_2^{\sf th} \\
  G_{\sf b_L}^{\sf side} & \psi > \psi_2^{\sf th}
\end{cases},
\end{align}
where $\psi_1^{\sf th}$ and $\psi_2^{\sf th}$ represent threshold elevation angles. Whereas, we model the antenna gain of BS to the satellite user as a two-lobe approximation with $G_{\sf b}^{\sf main}$ and $G_{\sf b}^{\sf side}=G_{\sf b_H}^{\sf side}$. This is because the impact of the BS antenna gain on satellite users is significant due to the downward tilt of BS antennas.

% To address this, we model the transmit side lobe beam gain of BS as a function of the distance between the BSs and the typical satellite $\norm{\mathbf{b}_m-\mathbf{x}_1}$, as follows:
% \begin{align}
%     G_{\sf b}^{\sf side} = f_{G_{\sf b}^{\sf side}}(\norm{\mathbf{b}_m-\mathbf{x}_1}),
% \end{align}
% where the distance between the BS $\mathbf{b}_m$ for the elevation angle $\psi_{m}$ and the typical satellite $\mathbf{x}_1$ is computed by
% \begin{align}
%     \norm{\mathbf{b}_m-\mathbf{x}_1}=\sqrt{R_{\sf s}^2-R_{\sf b}^2\cos^2{\psi_m}}-R_{\sf b}\sin{\psi_m}.
% \end{align}
% This approach enhances our understanding of the influence of the elevation angle on the side-lobe beam gain. The function $f_{G_{\sf b}^{\sf side}}(\norm{\mathbf{b}_m-\mathbf{x}_1})$ can be represented by polynomial or exponential functions. 

In UL satellite networks, we assume that the receive beam of the typical satellite is perfectly aligned with the transmit beam of the nearest satellite user $\mathbf{u}^{s}_1$. Further, we assume that the receive beam of the typical satellite is misaligned with the transmit beam of other transmitters. Then, the effective antenna gain of the typical satellite from the satellite user $\ell$ is modeled as
\begin{align}
    G_{\ell}^{\sf u_s|s} =\begin{cases}
  G_{\sf s}^{\sf main}G_{\sf u_s}^{\sf main}\left(\frac{c}{4\pi f_c}\right)^2  & \ell = 1 \\
  G_{\sf s}^{\sf side}G_{\sf u_s}^{\sf side}\left(\frac{c}{4\pi f_c}\right)^2 & \text{otherwise}
\end{cases},
\end{align}
where $f_c$ is the carrier frequency and $c$ is the speed of light.
Further, we assume that $\psi_1^{\sf th}<\cos^{-1}\left(\frac{R_{\sf s}}{R_{\sf s}}\sin \theta_{\sf s}\right)$, which means that the typical satellite is located within the side-lobe beam of the BSs. Then, the effective antenna gains of the typical satellite from the terrestrial users $(o={\sf u_t})$ and BSs $(o={\sf b})$ are also modeled as
\begin{align}
    G^{o|{\sf s}} = G_{\sf s}^{\sf side}G_{o}^{\sf side}\left(\frac{c}{4\pi f_c}\right)^2.
\end{align}

Similarly, in DL satellite networks, the receive beam of the typical satellite user is assumed to be perfectly aligned with the transmit beam of the nearest satellite $\mathbf{x}_1$. Further, we assume that the receive beam of the typical satellite user is misaligned with the transmit beam of other transmitters. Then, the effective antenna gain of the typical satellite user from the satellite $n$ is modeled as
\begin{align}
    G_{n}^{\sf s|u_s} =\begin{cases}
  G_{\sf u_s}^{\sf main}G_{\sf s}^{\sf main}\left(\frac{c}{4\pi f_c}\right)^2  & n = 1 \\
  G_{\sf u_s}^{\sf side}G_{\sf s}^{\sf side}\left(\frac{c}{4\pi f_c}\right)^2 & \text{otherwise}
\end{cases}.
\end{align}
Further, the effective antenna gains of the typical satellite user from the terrestrial user $(o={\sf u_t})$ and BSs $(o={\sf b_H})$ are also modeled as
\begin{align}
    G^{o|{\sf u_s}} = G_{\sf u_s}^{\sf side}G_{ o}^{\sf side}\left(\frac{c}{4\pi f_c}\right)^2.
\end{align}

\subsection{Channel Fading Model}
To model the fading effect of satellite channels, we adopt a Shadowed-Rician channel model. This fading distribution is used to characterize the channels between satellites and both satellite users and terrestrial nodes. Let $\sqrt{H_{i}^{\sf s}}\in \mathbb{C}$ be the complex channel coefficient of the satellite channel, then the probability density function (PDF) of $\sqrt{H_{i}^{\sf s}}$ is given by
\begin{align}
    f_{\sqrt{H_{i}^{\sf s}}}(x)&=\left(\frac{2b{m_{\sf s}}}{2b{m_{\sf s}}+\Omega}\right)^{m_{\sf s}} \frac{x}{b}\exp\left(\frac{-x^2}{2b}\right) \nonumber\\ &~~~~~~~~~~~~\cdot F_1\left({m_{\sf s}};1;\frac{\Omega x^2}{2b(2b{m_{\sf s}}+\Omega)}\right),
\end{align}
where $F_1(a;b;c)$ is the confluent hyper-geometric function of the first kind \cite{magnus1967}, $2b$ and $\Omega$ are the average power of the scatter component and line-of-sight (LOS) component, respectively, and ${m_{\sf s}}$ is the Nakagami parameter for the satellite channel.

Further, we model the fading effect of terrestrial channels as the Nakagami channel model. This fading distribution is used for modeling the channels between satellite users and terrestrial nodes. Let $\sqrt{H_{\ell}^{\sf t}}\in \mathbb{C}$ be the complex channel coefficient of the terrestrial channel. Then, PDF of $\sqrt{H_{\ell}^{\sf t}}$ is given by
\begin{align}
    f_{\sqrt{H_{\ell}^{\sf t}}}(x)=\frac{2{m_{\sf t}}^{m_{\sf t}}}{\Gamma({m_{\sf t}})}x^{2{m_{\sf t}}-1}\exp\left(-{m_{\sf t}}x^2\right),
\end{align}
where ${m_{\sf t}}$ is the Nakagami parameter for the terrestrial channel.

% Further, we use the Nakagami channel model for modeling the fading effect of terrestrial channels. The PDF of complex channel coefficient from the $j$th terrestrial user or BS to the typical satellite $\sqrt{H_j^{\sf t}}$ is given by
% \begin{align}
%     f_{\sqrt{H_{j}^{\sf t}}}(x)=\frac{2{m_{\sf t}}^{m_{\sf t}}}{\Gamma({m_{\sf t}})}x^{2{m_{\sf t}}-1}\exp\left(-{m_{\sf t}}x^2\right),
% \end{align}
% ${m_{\sf t}}$ is the Nakagami parameter for the terrestrial channel.

\subsection{Signal Model}
\subsubsection{UL Satellite Networks}
We first consider the UL satellite networks considering spectrum sharing with DL and UL terrestrial networks.
Let $P_{o}$ be the transmit power for $o\in\{\sf u_s,b,u_t\}$. Then, the SINR for the UL typical satellite served by the nearest satellite user when spectrum sharing with DL terrestrial networks is given by
\begin{align}
    {\sf SINR_{up}^{dn}} &= \frac{G_{1}^{\sf u_s|s} P_{\sf u_s} H_1^{\sf s} \norm{\mathbf{u}_1^{\sf s}-\mathbf{x}_1}^{-\alpha_{\sf s}}}{I_{\sf s}^{\sf up} +I_{\sf t,up}^{\sf dn} + \sigma^2},
\end{align}
where $I_{\sf s}^{\sf up}=\sum_{\mathbf{u}_{\ell}^{\sf s} \in \Phi_{\sf u_{\sf s}}\cap \mathcal{A}_{\sf u_{\sf s}|s}/\mathbf{u}_1^{\sf s}}G_{\ell}^{\sf u_s|s} P_{\sf u_s} H^{\sf s}_{\ell}\norm{\mathbf{u}_{\ell}^{\sf s}-\mathbf{x}_1}^{-\alpha_{\sf s}} $, $I_{\sf t,up}^{\sf dn}=\sum_{\mathbf{b}_{m}\in \Phi_{\sf b} \cap \mathcal{A}_{\sf b|s}}G^{\sf b|s} P_{\sf b} H_{m}^{\sf s} \norm{\mathbf{b}_{m}-\mathbf{x}_1}^{-\alpha_{\sf s}}$, $\sigma^2$ denotes the noise power, and $\alpha_{\sf s}$ is the path-loss exponent of satellite links.

On the other hand, the SINR for the UL typical satellite served by the satellite user when spectrum sharing with UL terrestrial networks is given by
\begin{align}
    {\sf SINR_{up}^{up}} &= \frac{G_{1}^{\sf u_s|s} P_{\sf u_s} H_1^{\sf s} \norm{\mathbf{u}_1^{\sf s}-\mathbf{x}_1}^{-\alpha_{\sf s}}}{I_{\sf s}^{\sf up} +I_{\sf t,up}^{\sf up} + \sigma^2},
\end{align}
where $I_{\sf t,up}^{\sf up}=\sum_{\mathbf{u}_{k}^{\sf t} \in \Phi_{\sf u_{\sf t}}\cap \mathcal{A}_{\sf u_{\sf t}|s}}G^{\sf u_t|s} P_{\sf u_t}  H^{\sf s}_{k} \norm{\mathbf{u}_{k}^{\sf t}-\mathbf{x}_1}^{-\alpha_{\sf s}} $.

\vspace{0.2cm}
\subsubsection{DL Satellite Networks}
We also consider the DL satellite networks considering the spectrum sharing with DL and UL terrestrial networks. Let $P_{\sf s}$ be the transmit power for the satellite. Then, the SINR for the DL typical satellite user served by the nearest satellite when spectrum sharing with DL terrestrial networks is given by
\begin{align}
    {\sf SINR_{dn}^{dn}} &= \frac{G_{1}^{\sf s|u_s} P_{\sf s} H_1^{\sf s} \norm{\mathbf{x}_1-\mathbf{u}_1^{\sf s}}^{-\alpha_{\sf s}}}{I_{\sf s}^{\sf dn} +I_{\sf t,dn}^{\sf dn} + \sigma^2},
\end{align}
where $I_{\sf s}^{\sf dn}=\sum_{\mathbf{x}_{n} \in \Phi_{\sf s}\cap \mathcal{A}_{\sf s|u_s}/\mathbf{x}_1}G_{n}^{\sf s|u_s} P_{\sf s} H^{\sf s}_{n}\norm{\mathbf{x}_{n}-\mathbf{u}_1^{\sf s}}^{-\alpha_{\sf s}} $, $I_{\sf t,dn}^{\sf dn}=\sum_{\mathbf{b}_{m}\in \Phi_{\sf b} \cap \mathcal{A}_{\sf b|u_s}}G^{\sf b|u_s} P_{\sf b} H_{m}^{\sf t} \norm{\mathbf{b}_{m}-\mathbf{u}_1^{\sf s}}^{-\alpha_{\sf t}}$, and $\alpha_{\sf t}$ denotes the path-loss exponent of terrestrial link.

On the other hand, the SINR for the DL typical satellite served by the nearest satellite when spectrum sharing with UL terrestrial networks is given by
\begin{align}
    {\sf SINR_{dn}^{up}} &= \frac{G_{1}^{\sf s|u_s} P_{\sf s} H_1^{\sf s} \norm{\mathbf{x}_1-\mathbf{u}_1^{\sf s}}^{-\alpha_{\sf s}}}{I_{\sf s}^{\sf dn} +I_{\sf t,dn}^{\sf up} + \sigma^2},
\end{align}
where $I_{\sf t,dn}^{\sf up}=\sum_{\mathbf{u}_{k}^{\sf t} \in \Phi_{\sf u_{\sf t}}\cap \mathcal{A}_{\sf u_{\sf t}|u_s}}G^{\sf u_t|u_s} P_{\sf u_t}  H^{\sf t}_{k} \norm{\mathbf{u}_{k}^{\sf t}-\mathbf{u}_1^{\sf s}}^{-\alpha_{\sf t}} $.

\subsection{Performance Metric}

\subsubsection{UL Satellite Networks} We characterize the ergodic spectral efficiency for the typical satellite $\mathbf{x}_1$ under spectrum sharing with DL terrestrial networks, which is given by
\begin{align}
    R_{\sf up}^{\sf dn}\!=\!\mathbb{E}\left[\log_2 \left(1\!+\!{\sf SINR_{up}^{dn}}\right)\right] \!=\!\int_{0}^{\infty}\frac{\log_2 e}{1+\gamma}\mathbb{P}\left[{\sf SINR_{up}^{dn}} \ge \gamma \right] \mathrm{d}\gamma, \label{eq:rate_updn}
\end{align}
and under spectrum sharing with UL terrestrial networks, which is obtained by
\begin{align}
    R_{\sf up}^{\sf up}\!=\!\mathbb{E}\left[\log_2 \left(1\!+\!{\sf SINR_{up}^{up}}\right)\right]\! =\!\int_{0}^{\infty}\frac{\log_2 e}{1+\gamma}\mathbb{P}\left[{\sf SINR_{up}^{up}} \ge \gamma \right] \mathrm{d}\gamma. \label{eq:rate_upup}
\end{align}

\subsubsection{DL Satellite Networks}
 We characterize the ergodic spectral efficiency for the typical satellite user $\mathbf{u}_1^{\sf s}$ under spectrum sharing with DL terrestrial networks, which is given by
\begin{align}
    R_{\sf dn}^{\sf dn}\!=\!\mathbb{E}\left[\log_2 \left(1\!+\!{\sf SINR_{dn}^{dn}}\right)\right] \!=\!\int_{0}^{\infty}\frac{\log_2 e}{1+\gamma}\mathbb{P}\left[{\sf SINR_{dn}^{dn}} \ge \gamma \right] \mathrm{d}\gamma, \label{eq:rate_dndn}
\end{align}
and under spectrum sharing with UL terrestrial networks, which is obtained by
\begin{align}
    R_{\sf dn}^{\sf up}\!=\!\mathbb{E}\left[\log_2 \left(1\!+\!{\sf SINR_{dn}^{up}}\right)\right] \!=\!\int_{0}^{\infty}\frac{\log_2 e}{1+\gamma}\mathbb{P}\left[{\sf SINR_{dn}^{up}} \ge \gamma \right] \mathrm{d}\gamma. \label{eq:rate_dnup}
\end{align}

\section{Preliminaries}
In this section, we introduce several key lemmas essential for deriving the analytical expressions for the coverage probabilities discussed in Sections IV and V. The lemmas introduced in this section establish a unified framework for network analysis across four different spectrum sharing scenarios.

\subsection{Lemmas for UL Satellite Network Analysis}
We adopt the geometric simplification for UL satellite networks from a $\mathbb{R}^3$ sphere to a $\mathbb{R}^2$ plane, as introduced by  \cite{Kim2023}. In the circular ring $\tilde{\mathcal{A}}{o|\sf s}$, which has inner and outer radii $R_{\sf min}^{o|\sf s}$ and $R_{\sf max}^{o|\sf s}$ respectively, nodes are distributed according to a homogeneous PPP with density $\tilde{\lambda}_{o|\sf s}$, where $o \in \{\sf b, u_s, u_t\}$.
\begin{lemma}[A Replacement Lemma]
    By letting $\tilde{\lambda}_{o|\sf s}=\lambda_{o} \frac{R_o}{R_{\sf s}}$ for $o\in\{\sf b, u_s, u_t\}$, the statistical property in the visible spherical cap $\mathcal{A}_{o|\sf s}$ is identical in the transformed circular ring $\tilde{\mathcal{A}}_{o|\sf s}$.
    \begin{proof}
        See the proof of Lemma 1 in \cite{Kim2023}.
    \end{proof}
\end{lemma}

In the following sections, we use representation of $\tilde{\lambda}_{o|\sf s}$, $|\tilde{\mathcal{A}}_{o|\sf s}|=\pi\left((R_{\sf max}^{o|\sf s})^2-(R_{\sf min}^{o|\sf s})^2\right)$, and $|\tilde{\mathcal{A}}_{o|\sf s}(r)|=\pi\left(r^2-(R_{\sf min}^{o|\sf s})^2\right)$ for $o\in\{\sf b, u_s, u_t\}$.

%\subsection{Statistical Properties}
We provide a set of lemmas to derive the exact expressions for the ergodic spectral efficiency.
\begin{lemma}[The nearest satellite user distance distribution] \label{lem:nearest_distance}
    The PDF of distance between the typical satellite $\mathbf{x}_1$ and the nearest satellite user $\mathbf{u}_1^{\sf s}$ is given by
    \begin{align}
        &f_{\norm{\mathbf{u}_1^{\sf s}-\mathbf{x}_1}|\Phi_{\sf u_{\sf s}}(\mathcal{A}_{\sf u_{\sf s}|\sf s})\ge 1}(r) \nonumber\\ &=\frac{2\tilde{\lambda}_{\sf u_{\sf s} |\sf s}\pi e^{\Tilde{\lambda}_{\sf u_{\sf s}|\sf s} \pi \left(R_{\sf min}^{\sf u_{\sf s}|\sf s}\right)^2} }{1-e^{-\tilde{\lambda}_{\sf u_{\sf s}|\sf s } \pi \left((R_{\sf max}^{\sf u_{\sf s}|\sf s})^2-(R_{\sf min}^{\sf u_{\sf s}|\sf s})^2\right)}}r e^{-\tilde{\lambda}_{\sf u_{\sf s}|\sf s } \pi r^2},
    \end{align}
     for $R_{\sf min}^{\sf u_{\sf s}|\sf s} \le r \le R_{\sf max}^{\sf u_{\sf s}|\sf s}$.
    \begin{proof}
    See appendix \ref{append:lem:nearest_distance}.
\end{proof}
\end{lemma}

    \begin{lemma}[Laplace transform of the aggregated UL interference power plus noise power for the spectrum sharing with DL terrestrial networks] \label{lem:Laplace_updn}
          We define the aggregated interference plus noise power for $\norm{\mathbf{u}_1^{\sf s}-\mathbf{x}_1}=r$  when UL satellite networks share the spectrum with the DL terrestrial networks as
              \begin{align}
        I_{\sf up}^{\sf dn}(r) &= \sum_{\mathbf{u}_{\ell}^{\sf s} \in \Phi_{\sf u_{\sf s}}\cap \mathcal{A}_{\sf u_{\sf s}|s}/ \mathcal{A}_{\sf u_{\sf s}|s}(r)}G_{\ell}^{\sf u_s|s} P_{\sf u_s} H^{\sf s}_{\ell}\norm{\mathbf{u}_{\ell}^{\sf s}-\mathbf{x}_1}^{-\alpha_{\sf s}} \nonumber\\ &+ \sum_{\mathbf{b}_{m}\in \Phi_{\sf b} \cap \mathcal{A}_{\sf b}}G^{\sf b|s} P_{\sf b} H_{m}^{\sf s} \norm{\mathbf{b}_{m}-\mathbf{x}_1}^{-\alpha_{\sf s}} + \sigma^2.
    \end{align}
    Then, the conditional Laplace transform for the aggregated interference power plus noise power is given by
    \begin{align}
        &\mathcal{L}_{I_{\sf up}^{\sf dn}(r)|\Phi_{\sf u_{\sf s}}\left(\mathcal{A}_{\sf u_{\sf s}|s}\right)\ge 1}(s) =\exp\Bigg( -s \sigma^2 \nonumber\\&{-2\pi \tilde{\lambda}_{\sf u_{ \sf s} |s} \int_{r}^{R_{\sf max}^{\sf u_{ \sf s}|s }}\left[1\!-\!\sum_{z=0}^{m_{\sf s}-1}\! \frac{\zeta(z)\Gamma(z+1)}{(\beta-c+sG_{\ell}^{\sf u_s|s} P_{\sf u_s} v^{-\alpha_{ \sf s}})^{z+1}}\right]v\mathrm{d}v}\nonumber\\ &{-2\pi \tilde{\lambda}_{\sf b|s}\! \int_{R_{\sf min}^{\sf b|s}}^{R_{\psi'}^{\sf b}}\!\!\left[1\!-\!\sum_{i=0}^{m_{\sf s}-1}\frac{\zeta(i)\Gamma(i+1)}{(\beta-c+s G^{\sf b_L|s} P_{\sf b} w^{-\alpha_{\sf s}})^{i+1}}\right]\!w\mathrm{d}w} \nonumber\\ &{-2\pi \tilde{\lambda}_{\sf b|s}\! \int_{R_{\psi'}^{\sf b|s}}^{R_{\sf max}^{\sf b}}\!\!\left[1\!-\!\sum_{i=0}^{m_{\sf s}-1}\frac{\zeta(i)\Gamma(i+1)}{(\beta-c+s G^{\sf b_H|s} P_{\sf b} u^{-\alpha_{\sf s}})^{i+1}}\right]\!u\mathrm{d}u} \!\Bigg),    \label{eq:lem:Laplace_updn}
    \end{align}
    where $\beta = \frac{1}{2b}$, $c=\frac{\Omega}{2b(2bm+\Omega)}$, $\zeta(z)=\left(\frac{2bm}{2bm+\Omega}\right)^m \frac{\beta (-1)^z (1-m)_z c^z}{(z!)^2}$, $(x)_z=x(x+1)\cdots(x+z-1)$ is the Pochhammer symbol \cite{gradshteyn2014}, $\Gamma(z)=(z-1)!$ is the Gamma function for positive integer $z$, and $R_{\psi'}^{\sf b|s}=\sqrt{R_{\sf s}^2-R_{\sf b}^2\cos^2{\psi_2^{\sf th}}}-R_{\sf b}\sin{\psi_2^{\sf th}}$.

    \begin{proof}
    See appendix \ref{append:lem:Laplace_updn}.
    \end{proof}
    \end{lemma}

\begin{lemma}[Laplace transform of the aggregated UL interference power plus noise power for the spectrum sharing with UL terrestrial networks] \label{lem:Laplace_upup}
          We define the aggregated interference plus noise power for $\norm{\mathbf{u}_1^{\sf s}-\mathbf{x}_1}=r$  when the UL satellite networks share the spectrum with UL terrestrial networks as 
              \begin{align}
        I_{\sf up}^{\sf up}(r) &= \sum_{\mathbf{u}_{\ell}^{\sf s} \in \Phi_{\sf u_{\sf s}}\cap \mathcal{A}_{\sf u_{\sf s}|s}/ \mathcal{A}_{\sf u_{\sf s}|s}(r)}G_{\ell}^{\sf u_s|s} P_{\sf u_s} H^{\sf s}_{\ell}\norm{\mathbf{u}_{\ell}^{\sf s}-\mathbf{x}_1}^{-\alpha_{\sf s}} \nonumber\\ &+ \sum_{\mathbf{u}_{k}^{\sf t} \in \Phi_{\sf u_{\sf t}}\cap \mathcal{A}_{\sf u_{\sf t}|s}}G^{\sf u_t|s} P_{\sf u_t}  H^{\sf s}_{k} \norm{\mathbf{u}_{k}^{\sf t}-\mathbf{x}_1}^{-\alpha_{\sf s}} + \sigma^2.
    \end{align}

    Then, the conditional Laplace transform for the aggregated
interference power plus noise power is given by
    \begin{align}
        &\mathcal{L}_{I_{\sf up}^{\sf up}(r)|\Phi_{\sf u_{\sf s}}\left(\mathcal{A}_{\sf u_{\sf s}|s}\right)\ge 1}(s) =\exp\Bigg( -s \sigma^2 \nonumber\\&{-2\pi \tilde{\lambda}_{\sf u_{ \sf s} |s} \int_{r}^{R_{\sf max}^{\sf u_{ \sf s}|s }}\left[1\!-\!\sum_{z=0}^{m_{\sf s}-1}\! \frac{\zeta(z)\Gamma(z+1)}{(\beta-c+sG_{\ell}^{\sf u_s|s} P_{\sf u_s} v^{-\alpha_{ \sf s}})^{z+1}}\right]v\mathrm{d}v}\nonumber\\ &{-2\pi \tilde{\lambda}_{\sf u_t|s}\! \int_{R_{\sf min}^{\sf u_t|s}}^{R_{\sf max}^{\sf u_t|s}}\!\!\left[1\!-\!\sum_{i=0}^{m_{\sf s}-1}\frac{\zeta(i)\Gamma(i+1)}{(\beta \!-\!c\!+\!s G^{\sf u_t|s} P_{\sf u_t} w^{-\alpha_{\sf s}})^{i+1}}\right]\!w\mathrm{d}w} \!\Bigg).
    \end{align}
    \begin{proof}
        The proof is omitted because it can be easily derived in a manner similar to Lemma \ref{lem:Laplace_updn}.
    \end{proof}
    \end{lemma}

\subsection{Lemmas for Satellite DL Analysis}
We exploit the replacement lemma in \cite{kang2023} that transforms the DL satellite networks' geometric representation from $\mathbb{R}^3$ sphere to $\mathbb{R}^2$ plane. In circular ring $\tilde{\mathcal{A}}_{o|\sf u_s}$ with inner and outer radii are $R_{\sf min}^{o|\sf u_s}$ and $R_{\sf max}^{o|\sf u_s}$, the users are distributed according to the homogeneous PPP with density $\tilde{\lambda}_{o|\sf u_s}$ for $o\in\{\sf s, b\}$.
\begin{lemma}[A Replacement Lemma]
    By letting $\tilde{\lambda}_{o|\sf u_s}=\lambda_{o} \frac{R_o}{R_{\sf u_s}}$ for $o\in\{\sf s, b\}$, the statistical property in the visible spherical cap $\mathcal{A}_{o|\sf u_s}$ is identical in the transformed circular ring $\tilde{\mathcal{A}}_{o|\sf u_s}$.
    \begin{proof}
        See Lemma 1 in \cite{Kim2023}.
    \end{proof}
\end{lemma}

In the following sections, we use representation of $\tilde{\lambda}_{o|\sf u_s}$, $|\tilde{\mathcal{A}}_{o|\sf u_s}|=\pi\left((R_{\sf max}^{o|\sf u_s})^2-(R_{\sf min}^{o|\sf u_s})^2\right)$, and $|\tilde{\mathcal{A}}_{o|\sf u_s}(r)|=\pi\left(r^2-(R_{\sf min}^{o|\sf u_s})^2\right)$ for $o\in\{\sf s, b\}$.

\begin{lemma}[The nearest satellite distance distribution] \label{lem:nearest_sat_distance}
    The PDF of distance between the typical satellite user $\mathbf{u}_1^{\sf s}$ and the nearest satellite $\mathbf{x}_1$ is given by
    \begin{align}
        &f_{\norm{\mathbf{x}_1-\mathbf{u}_1^{\sf s}}|\Phi_{\sf {\sf s}}(\mathcal{A}_{\sf {\sf s}|\sf s})\ge 1}(r) \nonumber\\ &=\frac{2\tilde{\lambda}_{\sf {\sf s} |\sf u_s}\pi e^{\Tilde{\lambda}_{\sf {\sf s}|\sf u_s} \pi \left(R_{\sf min}^{\sf {\sf s}|\sf u_s}\right)^2} }{1-e^{-\tilde{\lambda}_{\sf {\sf s}|\sf u_s } \pi \left((R_{\sf max}^{\sf {\sf s}|\sf u_s})^2-(R_{\sf min}^{\sf {\sf s}|\sf u_s})^2\right)}}r e^{-\tilde{\lambda}_{\sf {\sf s}|\sf u_s } \pi r^2}.
    \end{align}
    \begin{proof}
    The proof is straightforward from Lemma \ref{lem:nearest_distance}.
\end{proof}
\end{lemma}

    \begin{lemma}[Laplace transform of the aggregated DL interference power plus noise power for the spectrum sharing with DL terrestrial networks] \label{lem:Laplace_dndn}
          We define the aggregated interference plus noise power for $\norm{\mathbf{x}_1-\mathbf{u}_1^{\sf s}}=r$  when the DL satellite networks share the spectrum with DL terrestrial networks as
              \begin{align}
        I_{\sf dn}^{\sf dn}(r) &= \sum_{\mathbf{x}_{n} \in \Phi_{\sf s}\cap \mathcal{A}_{\sf s|u_s}/ \mathcal{A}_{\sf s|u_s}(r)}G_{n}^{\sf s|u_s} P_{\sf s} H^{\sf s}_{n}\norm{\mathbf{x}_{n}-\mathbf{u}_1^{\sf s}}^{-\alpha_{\sf s}} \nonumber\\ &+ \sum_{\mathbf{b}_{m}\in \Phi_{\sf b} \cap \mathcal{A}_{\sf b|u_s}}G^{\sf b|u_s} P_{\sf b} H_{m}^{\sf t} \norm{\mathbf{b}_{m}-\mathbf{u}_1^{\sf s}}^{-\alpha_{\sf t}} + \sigma^2.
    \end{align}
    Then, the conditional Laplace transform is given by
    \begin{align}
        &\mathcal{L}_{I_{\sf dn}^{\sf dn}(r)|\Phi_{\sf s}\left(\mathcal{A}_{\sf s|u_{\sf s}}\right)\ge 1}(s) =\exp\Bigg( -s \sigma^2 \nonumber\\&{-2\pi \tilde{\lambda}_{\sf s|u_s} \int_{r}^{R_{\sf max}^{\sf s|u_{ \sf s} }}\left[1-\sum_{z=0}^{m_{\sf s}-1}\! \frac{\zeta(z)\Gamma(z+1)}{(\beta-c+sG_{\ell}^{\sf s|u_s} P_{\sf s} v^{-\alpha_{ \sf s}})^{z+1}}\right]v\mathrm{d}v} \nonumber\\ & {-2\pi \tilde{\lambda}_{\sf b|u_s}\! \int_{R_{\sf min}^{\sf b|u_s}}^{R_{\sf max}^{\sf b|u_s}}\!\!\left[\!1\!-\!{\left(\!1\!+\!\frac{sG^{\sf b_H|u_s}P_{\sf b}w^{-\alpha_{\sf t}}}{m_{\sf t}}\right)^{-m_{\sf t}}}\right]\!w\mathrm{d}w}\!\Bigg).
    \end{align}

    \begin{proof}
        See appendix \ref{append:lem:Laplace_dndn}
    \end{proof}
    \end{lemma}

\begin{lemma}[Laplace transform of the aggregated DL interference power plus noise power for the spectrum sharing with UL terrestrial networks] \label{lem:Laplace_dnup}
          We define the aggregated interference plus noise power for $\norm{\mathbf{x}_1-\mathbf{u}_1^{\sf s}}=r$  when the UL satellite networks share the spectrum with DL terrestrial networks as
              \begin{align}
        &I_{\sf dn}^{\sf up}(r) = \sum_{\mathbf{x}_{n} \in \Phi_{\sf s}\cap \mathcal{A}_{\sf s|u_s}/ \mathcal{A}_{\sf s|u_s}(r)}G_{n}^{\sf s|u_s} P_{\sf s} H^{\sf s}_{n}\norm{\mathbf{x}_{n}-\mathbf{u}_1^{\sf s}}^{-\alpha_{\sf s}} \nonumber\\ &~~~+ \sum_{\mathbf{u}_{k}^{\sf t}\in \Phi_{\sf u_t} \cap \mathcal{A}_{\sf u_t|u_s}}G^{\sf u_t|u_s} P_{\sf u_t} H_{k}^{\sf t} \norm{\mathbf{u}_{k}^{\sf t}-\mathbf{u}_1^{\sf s}}^{-\alpha_{\sf t}} + \sigma^2.
    \end{align}
    Then, the conditional Laplace transform is given by
    \begin{align}
        &\mathcal{L}_{I_{\sf dn}^{\sf up}(r)|\Phi_{\sf s}\left(\mathcal{A}_{\sf s|u_{\sf s}}\right)\ge 1}(s) =\exp\Bigg( -s \sigma^2 \nonumber\\&{-2\pi \tilde{\lambda}_{\sf s|u_s} \int_{r}^{R_{\sf max}^{\sf s|u_{ \sf s} }}\left[1\!-\!\sum_{z=0}^{m_{\sf s}-1}\! \frac{\zeta(z)\Gamma(z+1)}{(\beta-c+sG_{\ell}^{\sf s|u_s} P_{\sf s} v^{-\alpha_{ \sf s}})^{z+1}}\right]v\mathrm{d}v} \nonumber\\ & {-2\pi {\lambda}_{\sf u_t}\! \int_{0}^{R_{\sf max}^{\sf u_t|u_s}}\!\!\left[\!1\!-\!{\left(\!1\!+\!\frac{sG^{\sf u_t|u_s}P_{\sf u_t}w^{-\alpha_{\sf t}}}{m_{\sf t}}\right)^{-m_{\sf t}}}\right]\!w\mathrm{d}w}\!\Bigg).
    \end{align}

    \begin{proof}
        The proof is omitted because it can be easily derived in a manner similar to Lemma \ref{lem:Laplace_dndn}.
    \end{proof}
    \end{lemma}

We are now prepared to present the ergodic spectral efficiency expressions for four spectrum sharing scenarios. Specifically, we will focus on analyzing the performance of UL and DL satellite networks under each of the two distinct spectrum sharing cases with terrestrial networks. Utilizing the lemmas derived earlier, one can readily derive the ergodic spectral efficiency expressions for both the UL and DL terrestrial networks under all possible spectrum sharing scenarios. However, due to page limitations, these cases will be omitted.

\section{Satellite UL Analysis }
In this section, we present expressions for the ergodic spectral efficiency of UL satellite networks under two distinct spectrum sharing scenarios with terrestrial networks.

\subsection{Exact Expression}
The following theorem presents the exact expression for the ergodic spectral efficiency of UL satellite networks in two spectrum sharing scenarios.

\begin{theorem}  \label{theorem:SE_up}
    The ergodic spectral efficiency for UL satellite networks when sharing spectrum with DL $(o={\sf dn})$ or UL $(o={\sf up})$  terrestrial networks is obtained by 
\begin{align}
        &R_{\sf up}^{o}= \left(1-e^{-\Tilde{\lambda}_{\sf u_s|\sf s} \pi \left((R_{\sf max}^{\sf u_s|\sf s})^2-(R_{\sf min}^{\sf u_s|\sf s})^2\right)}\right)\nonumber\\& \cdot\int_0^{\infty}\!\!\int_{R_{\sf min}^{\sf u_s|s}}^{R_{\sf max}^{\sf u_s|s}} \!\!\vast[\!1\!-\!\!\sum_{z=0}^{m_{\sf s}-1}\frac{\zeta(z)\Gamma(z)}{(\beta\!-\!c)^{z+1}}\vast(\!1\! -\!\sum_{v=0}^{z} \frac{((\beta-c)\gamma r^{\alpha_{\sf s}})^v}{\left(G_{1}^{\sf u_s|s} P_{\sf u_s}\right)^{v}v!}\nonumber\\&~~~~~~\cdot(-1)^v\frac{\mathrm{d}^v \mathcal{L}_{I_{\sf up}^{o}(r)|\Phi_{\sf u_s}(\mathcal{A}_{\sf u_s|s})\ge 1}(s)}{\mathrm{d}s^v} \bigg|_{ s= \frac{(\beta-c)\gamma r^{\alpha_{\sf s}}}{G_{1}^{\sf u_s|s} P_{\sf u_s}}  } \vast)\vast] \nonumber\\ &~~~~~~~~~~~\cdot f_{\norm{\mathbf{u}_1^{\sf s}-\mathbf{x}_1}|\Phi_{\sf u_{\sf s}}(\mathcal{A}_{\sf u_{\sf s}|s})\ge 1}(r) \frac{\log_2{e}}{1+\gamma} \mathrm{d}r \mathrm{d}\gamma.    \label{eq:theorem1}
    \end{align}
    \begin{proof}
            See appendix \ref{append:theorem:SE_up}.
    \end{proof}
\end{theorem}

\setlength{\tabcolsep}{3pt}
\begin{table}[!t]
\caption{Simulation Parameters} \label{Table:param}
\centering
\begin{tabular}{|ccc|}
\hline
\multicolumn{3}{|c|}{Satellite network parameters}\\ \hline  
\multicolumn{1}{|c|}{Satelliete user type} & \multicolumn{1}{c|}{VSAT} & Handheld \\ \hline
\multicolumn{1}{|c|}{Carrier frequency $f_c$ [GHz]} & \multicolumn{1}{c|}{28} &  1.99 \\ \hline
\multicolumn{1}{|c|}{UL and DL bandwidth $W$ [MHz]} & \multicolumn{1}{c|}{50} & 5 \\ \hline
\multicolumn{1}{|c|}{Transmit power of satellite $P_{\sf s}$ [dBm]} & \multicolumn{2}{c|}{43} \\ \hline
\multicolumn{1}{|c|}{Transmit power of satellite user $P_{\sf u_s}$ [dBm]} & \multicolumn{1}{c|}{35.05} & 23 \\ \hline
\multicolumn{1}{|c|}{Main-lobe antenna gain of satellite $G_{\sf s}^{\sf main}$ [dBi]} & \multicolumn{1}{c|}{44.5} & 50 \\ \hline
\multicolumn{1}{|c|}{Side-lobe antenna gain of satellite $G_{\sf s}^{\sf side}$ [dBi]} & \multicolumn{1}{c|}{31.5} & 30 \\ \hline
\multicolumn{1}{|c|}{Main-lobe antenna gain of satellite user $G_{\sf u_s}^{\sf main}$ [dBi]} & \multicolumn{1}{c|}{34.2} & 0 \\ \hline
\multicolumn{1}{|c|}{Side-lobe antenna gain of satellite user $G_{\sf u_s}^{\sf side}$ [dBi]} & \multicolumn{1}{c|}{21.2} & 0 \\ \hline
\multicolumn{1}{|c|}{Radius of the Earth $R_{\sf E}$ [km]} & \multicolumn{2}{c|}{6,378} \\ \hline
\multicolumn{1}{|c|}{Satellite altitude $h_{\sf s}=R_{\sf s}-R_{\sf E}$ [km]} & \multicolumn{2}{c|}{530} \\ \hline
\multicolumn{1}{|c|}{Satellite user altitude  $h_{\sf u_s}=R_{\sf u_s}-R_{\sf E}$ [m]} & \multicolumn{2}{c|}{1.5} \\ \hline
\multicolumn{1}{|c|}{Visible angle of the typical satellite $\theta_{\sf s}$ [degree]} & \multicolumn{2}{c|}{57} \\ \hline
\multicolumn{1}{|c|}{Elevation angle of the typical satellite user $\theta_{\sf u_s}$ [degree]} & \multicolumn{2}{c|}{0} \\ \hline
\multicolumn{1}{|c|}{Path-loss exponent $\alpha_{\sf s}$} & \multicolumn{2}{c|}{2} \\ \hline
\multicolumn{1}{|c|}{Noise spectral density $N_0$ [dBm/Hz]} & \multicolumn{2}{c|}{-174} \\ \hline
\hline 
\multicolumn{3}{|c|}{Terrestrial network parameters} \\   \hline
\multicolumn{1}{|c|}{Trnasmit power of BS $P_{\sf b}$ [dBm]} & \multicolumn{2}{c|}{46} \\ \hline
\multicolumn{1}{|c|}{Trnasmit power of terrestrial user $P_{\sf u_t}$ [dBm]} & \multicolumn{2}{c|}{23} \\ \hline
\multicolumn{1}{|c|}{Side-lobe antenna gain of BS $G_{\sf b_H}^{\sf side}$, $G_{\sf b_L}^{\sf side}$ [dBi]} & \multicolumn{2}{c|}{4, -12} \\ \hline
\multicolumn{1}{|c|}{Side-lobe antenna gain of terrestrial user $G_{\sf u_t}^{\sf side}$ [dBi]} & \multicolumn{2}{c|}{0} \\ \hline
\multicolumn{1}{|c|}{Threshold elevation angle $\psi'$ [degree]} & \multicolumn{2}{c|}{40} \\ \hline
\multicolumn{1}{|c|}{BS altitude  $h_{\sf b}=R_{\sf b}-R_{\sf E}$ [m]} & \multicolumn{2}{c|}{35} \\ \hline
\multicolumn{1}{|c|}{Terrestrial user altitude  $h_{\sf u_t}=R_{\sf u_t}-R_{\sf E}$ [m]} & \multicolumn{2}{c|}{1.5} \\ \hline
\multicolumn{1}{|c|}{Path-loss exponent $\alpha_{\sf t}$} & \multicolumn{2}{c|}{4} \\ \hline
\end{tabular}
\end{table}

%%%%%%%%%%%%%%%%%%%%%%%%%%%%%%%%%%%%%%%%%%%%%%%%%%%%%%%%%%%%%%%%%%%%%%

\begin{figure}[t]
    \centering
    \subfloat[]{\includegraphics[width=9.5cm]{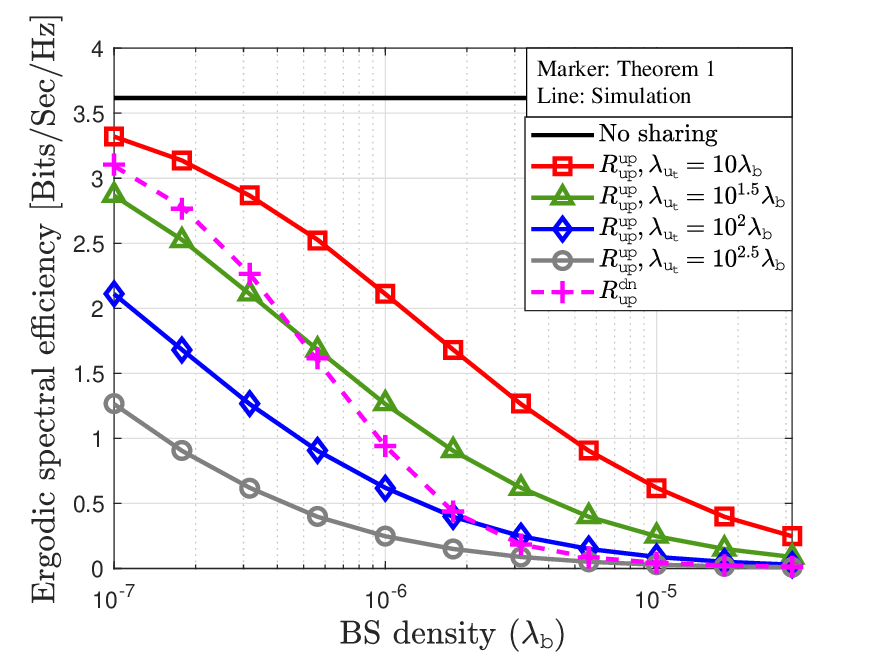}}\\ \vspace{-11.5pt}
    \subfloat[]{\includegraphics[width=9.5cm]{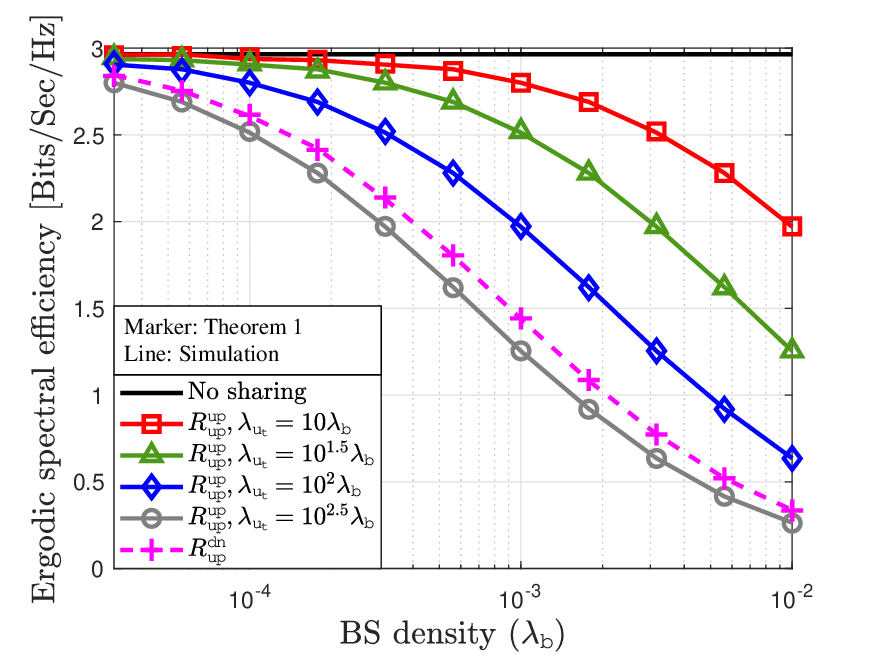}}
    \caption{The UL ergodic spectral efficiencies for the typical satellite with (a) handheld satellite users with $\lambda_{\sf u_s}=10^{-4.5}$ and  (b) VSAT satellite users with $\lambda_{\sf u_s}=10^{-5.5}$,  under spectrum sharing with terrestrial networks when $[m, b, \Omega]=[1,0.063,8.97 \times 10^{-4}]$.} \label{fig:UL_SE}
\end{figure}

Fig. \ref{fig:UL_SE} illustrates the UL ergodic spectral efficiencies when the spectrum of UL satellite networks is shared with terrestrial networks. The parameters used in the simulations are detailed in Table \ref{Table:param}, with references drawn from \cite{FCC91A1, 3GPP,FCCstardish}. In the figure, dashed lines with markers depict the spectral efficiency results for spectrum sharing with DL terrestrial networks, while solid lines with markers represent those for spectrum sharing with UL terrestrial networks. A comparison between Fig. \ref{fig:UL_SE}-(a) and Fig. \ref{fig:UL_SE}-(b) demonstrates that spectrum sharing in a handheld scenario leads to a significant reduction in ergodic spectral efficiency. This pronounced degradation is primarily due to the dominant interference from VSAT users in the VSAT scenario, which exceeds that from  BSs and terrestrial users. Conversely, in the handheld scenario, the interference from handheld users is considerably lower due to their reduced transmit power and antenna gains, making the interference from BSs and terrestrial users more pronounced and adversely affecting the ergodic spectral efficiency.

Furthermore, a comparison of the ergodic spectral efficiencies in the two spectrum sharing scenarios shown in Fig. \ref{fig:UL_SE}-(a) indicates that spectrum sharing with DL terrestrial networks results in higher ergodic spectral efficiencies than sharing with UL terrestrial networks when terrestrial user densities are more than $10^{2}$ times the BS density. Similarly, in Fig. \ref{fig:UL_SE}-(b), the ergodic spectral efficiencies under spectrum sharing with DL terrestrial networks are higher than those with UL terrestrial networks when terrestrial user densities exceed $10^{2.5}$ times the BS density. 

% Given that the number of users per BS typically exceeds $10^{2.5}$ in LTE and 5G networks, spectrum sharing with DL terrestrial networks is generally more beneficial, despite the higher transmit power and antenna gain of BSs compared to terrestrial users. This advantage arises because BS antennas are down-tilted, reducing the impact of their antenna gain on the satellite.

\subsection{Special Cases}
The exact expression for the ergodic spectral efficiency of the UL satellite network, as derived in Theorem 1, fully encapsulates all relevant network and system parameters. However, this comprehensive formulation makes it challenging to discern the interactions among key network and system parameters. To gain clearer insights, we simplify the original expression by examining specific special cases. Within these simplified scenarios, we will specifically explore how changes in various system parameters influence the satellite UL spectral efficiency.

We examine a special case where the path-loss exponent is $\alpha_{\sf s} = 2$ and the channel fading distributions follow Rayleigh fading, to elucidate the derived expressions and assess which spectrum sharing scenario offers more advantages given specific system parameters. In this scenario, we calculate the density ratio of terrestrial networks that makes spectrum sharing with UL terrestrial networks more beneficial than with DL terrestrial networks, specifically in terms of the lower bound of the ergodic spectral efficiency. The following corollary identifies a critical density ratio between DL terrestrial BSs  and UL terrestrial users, indicating when spectrum sharing with UL terrestrial networks is advantageous for UL satellite networks.

\begin{corollary}[UL-satellite and UL-terrestrial spectrum sharing condition] \label{cor:LB_UL}
For a path-loss exponent of \(\alpha_{\sf s} = 2\) and Rayleigh fading in the channel fading distributions, the lower bound of the ergodic spectral efficiency under spectrum sharing with DL terrestrial networks is less than that under spectrum sharing with UL terrestrial networks, provided the following density ratio is met:    \begin{align}
    \frac{\lambda_{\sf u_t}}{\lambda_{\sf b}}\le\frac{P_{\sf b}R_{\sf b}}{P_{\sf u_t}R_{\sf u_t}}\frac{G^{\sf b_L|s}\ln\left(\frac{R_{\psi'}^{\sf b|s}}{R_{\sf min}^{\sf b|s}}\right)+G^{\sf b_H|s}\ln\left(\frac{R_{\sf max}^{\sf b|s}}{R_{\psi'}^{\sf b|s}}\right)}{G^{\sf u_t|s}\ln\left(\frac{R_{\sf max}^{\sf u_t|s}}{R_{\sf min}^{\sf u_t|s}}\right)}. \label{eq:LB_UL}
\end{align}

    \begin{proof}
        See appendix \ref{append:cor:LB_UL}
    \end{proof}
\end{corollary}

According to Corollary \ref{cor:LB_UL}, the density ratio of terrestrial networks that favors a particular transmission mode depends on the ratios of transmit powers, network altitudes, and effective antenna gains to the distances from the typical satellite user. Furthermore, using the network parameters listed in Table \ref{Table:param}, the condition specified in \eqref{eq:LB_UL} suggests that spectrum sharing with UL terrestrial networks is advantageous when \(\frac{\lambda_{\sf u_t}}{\lambda_{\sf b}} \leq 235\). This conclusion aligns closely with the simulation results depicted in Fig. \ref{fig:UL_SE}.

\section{Satellite DL Analysis}
%\subsection{A Replacement Lemma}

In this section, we provide analytical expressions for the ergodic spectral efficiency of DL satellite networks under two distinct spectrum sharing scenarios with terrestrial networks.

\subsection{Exact Expression}
The following theorem presents the exact expression for the ergodic spectral efficiency of DL satellite networks in two spectrum sharing scenarios.

\begin{theorem}  \label{theorem:SE_dn}
    The ergodic spectral efficiency for DL satellite networks when sharing spectrum with DL $(o={\sf dn})$ or UL $(o={\sf up})$ terrestrial networks is obtained by
    \begin{align}
        &R_{\sf dn}^o=\left(1-e^{-\Tilde{\lambda}_{\sf s|\sf u_s} \pi \left((R_{\sf max}^{\sf s|\sf u_s})^2-(R_{\sf min}^{\sf s|\sf u_s})^2\right)}\right)\nonumber\\ &\cdot \int_0^{\infty}\!\! \int_{R_{\sf min}^{\sf s|u_s}}^{R_{\sf max}^{\sf s|\sf u_s}} \!\vast[\!1\!-\!\sum_{z=0}^{m_{\sf s}-1}\frac{\zeta(z)\Gamma(z)}{(\beta\!-\!c)^{z+1}}\!\vast(\!1\! -\!\sum_{v=0}^{z} \frac{((\beta \!-\!c)\gamma r^{\alpha_{\sf s}})^v}{\left(G_{1}^{\sf s|u_s} P_{\sf s}\right)^{v}v!}\nonumber\\&~~~~~~\cdot(-1)^v\frac{\mathrm{d}^v \mathcal{L}_{I_{\sf dn}^{o}(r)|\Phi_{\sf s}(\mathcal{A}_{\sf s|u_s})\ge 1}(s)}{\mathrm{d}s^v} \bigg|_{ s= \frac{(\beta-c)\gamma r^{\alpha_{\sf s}}}{G_{1}^{\sf s|u_s} P_{\sf s}}  } \vast)\vast] \nonumber\\ &~~~~~~~~~~~~~~~~~~~\cdot f_{\norm{\mathbf{x}_1-\mathbf{u}_1^{\sf s}}|\Phi_{\sf s}(\mathcal{A}_{\sf s|u_s})\ge 1}(r) \frac{\log_2{e}}{1+\gamma}\mathrm{d}r \mathrm{d}\gamma.  \label{eq:theorem:SE_dn}
    \end{align}
    \begin{proof}
        See appendix \ref{append:theorem:SE_dn}.
    \end{proof}
\end{theorem}

\begin{figure}[t]
    \centering
    \subfloat[]{\includegraphics[width=9.5cm]{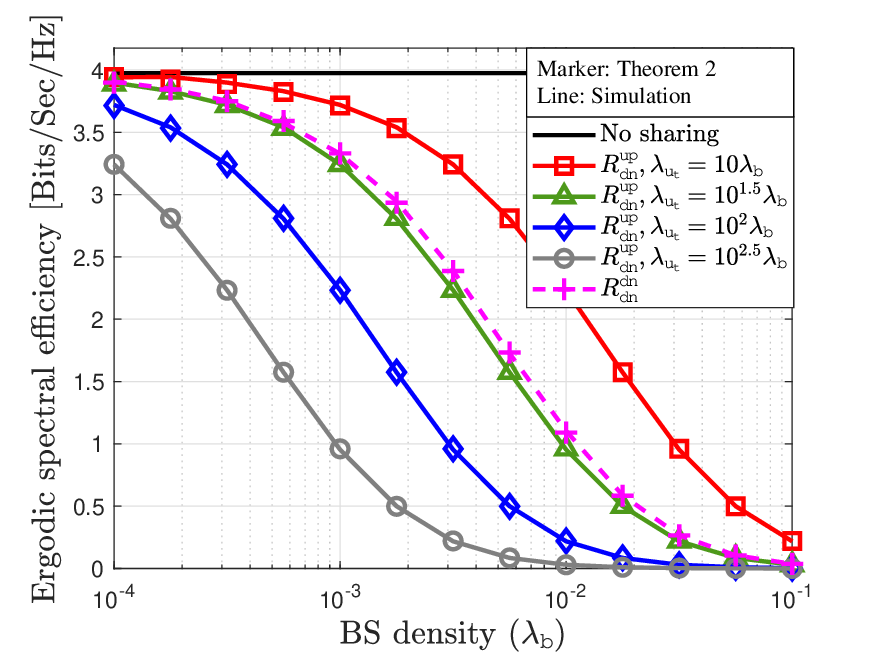}}\\ \vspace{-11.5pt}
    \subfloat[]{\includegraphics[width=9.5cm]{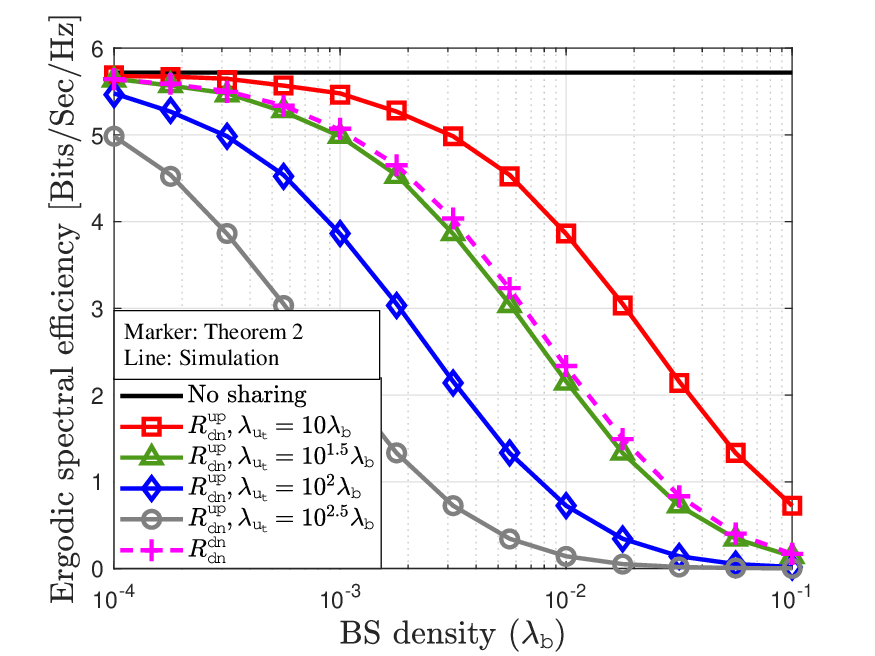}}
    \caption{The DL ergodic spectral efficiencies for the typical (a) handheld satellite user and  (b) VSAT satellite user under spectrum sharing with terrestrial networks when $[m, b, \Omega]=[1,0.063,8.97 \times 10^{-4}]$ and $\lambda_{\sf s}=10^{-6}$.} \label{fig:DL_SE}
\end{figure}

Fig. \ref{fig:DL_SE} shows the DL ergodic spectral efficiencies when the spectrum of DL satellite networks is shared with terrestrial networks. The dashed line with markers illustrates the results of spectrum sharing with DL terrestrial networks, while the solid lines with markers represent the results of sharing with uplink (UL) terrestrial networks. Both Fig. \ref{fig:DL_SE}-(a) and Fig. \ref{fig:DL_SE}-(b) show that spectrum sharing with DL terrestrial networks yields higher spectral efficiencies compared to sharing with UL terrestrial networks when terrestrial user densities exceed $10^{1.5}$ times the BS density. Considering that the number of users per BS typically surpasses $10^{1.5}$ in LTE and 5G networks, sharing with DL terrestrial networks generally offers more benefits. This advantage is primarily due to the increased likelihood of terrestrial users being in close proximity to satellite users as their density increases, despite the higher transmit power and antenna gain of BSs compared to terrestrial users

\subsection{Special Case}
 Similar to the analysis of UL spectral efficiency, it is beneficial to examine a special case in the DL spectral efficiency analysis, where the path-loss exponents are \(\alpha_{\sf s}=2\) for satellite networks and \(\alpha_{\sf t}=4\) for terrestrial networks, with Rayleigh fading for the channel fading distributions. Under these conditions, the DL ergodic spectral efficiency derived in Theorem 2 simplifies, providing valuable insights for network design.

 The following corollary determines the critical density ratio of terrestrial networks at which spectrum sharing with UL terrestrial networks becomes more advantageous than with DL terrestrial networks, as inferred from the lower bound of the ergodic spectral efficiency.

\begin{corollary}[DL-satellite and UL-terrestrial spectrum sharing condition] \label{cor:LB_DL}

For \(\alpha_{\sf s}=2\), \(\alpha_{\sf t}=4\), and Rayleigh fading in channel fading distributions, the lower bound of the ergodic spectral efficiency when spectrum sharing with DL terrestrial networks is less than that when sharing with UL terrestrial networks, provided the following density ratio is met:
    \begin{align}
    \frac{\lambda_{\sf u_t}}{\lambda_{\sf b}}\le\frac{P_{\sf b}R_{\sf b}G^{\sf b_H|s}}{P_{\sf u_t}R_{\sf u_s}G^{\sf u_t|u_s}}\frac{\frac{1}{\left(R_{\sf min}^{\sf b|u_s}\right)^2}-\frac{1}{\left(R_{\sf max}^{\sf b|u_s}\right)^2}}{\frac{1}{\left(R_{\sf min}^{\sf u_t|u_s}\right)^2}-\frac{1}{\left(R_{\sf max}^{\sf u_t|u_s}\right)^2}}. \label{eq:LB_DL}
\end{align}

    \begin{proof}
        See appendix \ref{append:cor:LB_DL}.
    \end{proof}
\end{corollary}

Using the network parameters listed in Table \ref{Table:param} and setting \(R_{\sf min}^{\sf u_t|u_s}= 7\) m, the condition specified in \eqref{eq:LB_DL} suggests that spectrum sharing with UL  terrestrial networks is advantageous when \(\frac{\lambda_{\sf u_t}}{\lambda_{\sf b}} \leq 35\). This finding is consistent with the simulation results depicted in Fig. \ref{fig:DL_SE}.

\section{Conclusion}
In this paper, we have investigated the ergodic spectral efficiency performance of UL and DL satellite networks considering spectrum sharing with terrestrial networks. Using stochastic geometry tools, we first derived the aggregated interference power from terrestrial and satellite nodes to the typical satellite node. Then, we derived the exact expressions for ergodic spectral efficiency for the UL and DL satellite networks in terms of network densities, fading parameters, and the path-loss exponent. Using simulations, we confirmed the accuracy of the derived expressions and compared the performance of two different spectrum-sharing configurations for each UL and DL satellite network. Our findings indicate that the ideal terrestrial transmission mode depends on the density ratio of terrestrial nodes. Furthermore, under the simplified fading distribution assumption, we determined the density ratio of terrestrial nodes that dictates the advantageous transmission configuration of terrestrial networks as a function of relevant network parameters. 

An intriguing direction for future research is to analyze the impact of advanced interference management techniques on spectrum sharing, including the use of coordinated beamforming \cite{Lee2016mag,Kim2023,JT,JT2}. Another promising research direction is to incorporate various stochastic models in the analysis of terrestrial and satellite locations, aiming to capture the orbital planes and repulsion effects among the nearest points \cite{Choi1,Choi2,Lee2022}.

\begin{appendices}
\section{Proof of Lemma \ref{lem:nearest_distance}}
\begin{proof}
First, we compute the probability of the number of visible satellite users is over $1$ as
    \begin{align}
        \mathbb{P}[\Phi_{\sf u_{\sf s}}(\mathcal{A}_{\sf u_{\sf s}|\sf s})\ge 1] &= 1-\mathbb{P}[\Phi_{\sf u_{\sf s}}(\mathcal{A}_{\sf u_{\sf s}|\sf s})= 0] \nonumber\\ &= 1-e^{-\left(\Tilde{\lambda}_{\sf u_s|\sf s}\big|\Tilde{\mathcal{A}}_{\sf u_{\sf s}|\sf s}\big|\right)} \nonumber \\ &= 1-e^{-\Tilde{\lambda}_{\sf u_s|\sf s} \pi \left((R_{\sf max}^{\sf u_s|\sf s})^2-(R_{\sf min}^{\sf u_s|\sf s})^2\right)}.
    \end{align}
    To compute the $f_{\norm{\mathbf{u}_1^{\sf s}-\mathbf{x}_1}|\Phi_{\sf u_{\sf s}}(\mathcal{A}_{\sf u_{\sf s}|\sf s})\ge 1}(r)$, we compute the probability that the nearest satellite user distance is larger than $r$ conditioned on $\Phi(\mathcal{A}_{\sf u_{\sf s}|\sf s})\ge 1$ as
    \begin{align}
        &\mathbb{P}\left[\norm{\mathbf{u}_1^{\sf s}-\mathbf{x}_1}>r|\Phi_{\sf u_{\sf s}}(\mathcal{A}_{\sf u_{\sf s}|\sf s})\ge 1\right] \nonumber\\ &= \mathbb{P}\left[\Phi_{\sf u_{\sf s}}\left(\mathcal{A}_{\sf u_{\sf s}|s}(r)\right)=0|\Phi_{\sf u_{\sf s}}(\mathcal{A}_{\sf u_{\sf s}|s})\ge 1\right] \nonumber\\ &=\frac{\mathbb{P}\left[\Phi_{\sf u_{\sf s}}\left(\mathcal{A}_{\sf u_{\sf s}|s}(r)\right)=0,\Phi_{\sf u_{\sf s}}(\mathcal{A}_{\sf u_{\sf s}|s})\ge 1\right]}{\mathbb{P}\left[\Phi_{\sf u_{\sf s}}(\mathcal{A}_{\sf u_{\sf s}|s})\ge 1\right]} \nonumber\\&\overset{(a)}{=} \frac{\mathbb{P}[\Phi_{\sf u_{\sf s}}\left(\mathcal{A}_{\sf u_{\sf s}|s}(r)\right)=0]\mathbb{P}[\Phi_{\sf u_{\sf s}}(\mathcal{A}_{\sf u_{\sf s}|s}/\mathcal{A}_{\sf u_{\sf s}|s}(r))\ge 1]}{\mathbb{P}\left[\Phi_{\sf u_{\sf s}}(\mathcal{A}_{\sf u_{\sf s}|s})\ge 1\right]} \nonumber\\&=\frac{e^{-\left(\tilde{\lambda}_{\sf u_{\sf s} |s}\big|\Tilde{\mathcal{A}}_{\sf u_{\sf s}|s}(r)\big|\right)}\left( 1- e^{-\left(\tilde{\lambda}_{\sf u_{\sf s} |s}\big|\Tilde{\mathcal{A}}_{\sf u_{\sf s}|s}/\Tilde{\mathcal{A}}_{\sf u_{\sf s}|s}(r)\big|\right)}\right)}{1-e^{-\left(\tilde{\lambda}_{\sf u_{\sf s} |s}\big|\Tilde{\mathcal{A}}_{\sf u_{\sf s}|s}\big|\right)}} \nonumber\\
        &= \frac{e^{-\tilde{\lambda}_{\sf u_{\sf s} |s} \pi \left(r^2-(R_{\sf min}^{\sf u_{\sf s} |s})^2\right)}-e^{-\tilde{\lambda}_{\sf u_{\sf s} |s} \pi \left((R_{\sf max}^{\sf u_{\sf s}|s })^2-(R_{\sf min}^{\sf u_{\sf s}|s })^2\right)}}{1-e^{-\tilde{\lambda}_{\sf u_{\sf s} |s} \pi \left((R_{\sf max}^{\sf u_{\sf s}|s})^2-(R_{\sf min}^{\sf u_{\sf s}|s})^2\right)}} \nonumber\\ &= F^c_{\norm{\mathbf{u}_1^{\sf s}-\mathbf{x}_1}|\Phi_{\sf u_{\sf s}}(\mathcal{A}_{\sf u_{\sf s}|s})\ge 1}(r),
    \end{align}
    where (a) follows from the independence of the PPP for non-overlapping areas $\mathcal{A}_{\sf u_{\sf s}|s}(r)$ and $\mathcal{A}_{\sf u_{\sf s}|s}/\mathcal{A}_{\sf u_{\sf s}|s}(r)$.
    By taking the derivative with respect to $r$, we obtain the nearest satellite user distance distribution as
    \begin{align}
        &f_{\norm{\mathbf{u}_1^{\sf s}-\mathbf{x}_1}|\Phi_{\sf u_{\sf s}}(\mathcal{A}_{\sf u_{\sf s}|\sf s})\ge 1}(r) =\frac{\partial(1-F^c_{\norm{\mathbf{u}_1^{\sf s}-\mathbf{x}_1}|\Phi_{\sf u_{\sf s}}(\mathcal{A}_{\sf u_{\sf s}|s})\ge 1}(r))}{\partial r}\nonumber\\ &=\frac{2\tilde{\lambda}_{\sf u_{\sf s} |\sf s}\pi e^{\Tilde{\lambda}_{\sf u_{\sf s}|\sf s} \pi \left(R_{\sf min}^{\sf u_{\sf s}|\sf s}\right)^2} }{1-e^{-\tilde{\lambda}_{\sf u_{\sf s}|\sf s } \pi \left((R_{\sf max}^{\sf u_{\sf s}|\sf s})^2-(R_{\sf min}^{\sf u_{\sf s}|\sf s})^2\right)}}r e^{-\tilde{\lambda}_{\sf u_{\sf s}|\sf s } \pi r^2},
    \end{align}
    for $R_{\sf min}^{\sf u_{\sf s}|\sf s} \le r \le R_{\sf max}^{\sf u_{\sf s}|\sf s}$.
\label{append:lem:nearest_distance}
\end{proof}

\section{Proof of Lemma \ref{lem:Laplace_updn}}
\begin{proof}
Let $\mathcal{A}^c_{\sf u_{ \sf s}|s}(r)=\mathcal{A}_{\sf u_{ \sf s}|s}/\mathcal{A}_{\sf u_{ \sf s}|s}(r)$. Then, the conditional Laplace transform for the aggregated interference power plus noise power is computed as
    \begin{align}
        &\mathcal{L}_{I_{\sf up}^{\sf dn}(r)|\Phi_{\sf u_{ \sf s}}\left(\mathcal{A}_{\sf u_{ \sf s}|s}\right)\ge 1}(s) \nonumber\\&= \mathbb{E}_{I_{\sf up}^{\sf dn}(r)}\!\left[\exp(-sI_{\sf up}^{\sf dn}(r))\big|\Phi_{\sf u_{ \sf s}}\!\left(\mathcal{A}_{\sf u_{ \sf s}|s}\right)\!\ge\! 1\right] 
        % \nonumber\\&= \mathbb{E}_{I_{\sf up}^{\sf dn}(r)}\!\Bigg[\!\exp\!\Bigg(\!\!\!-\!s\bigg( \sum_{\mathbf{u}_{\ell}^{ \sf s} \in \Phi_{\sf u_s}\cap \mathcal{A}^c_{\sf u_{ \sf s}|s}(r)}\!\!G_{\ell}^{\sf u_s|s} P_{\sf u_s}  H^{ \sf s}_{\ell} \norm{\mathbf{u}_{\ell}^{ \sf s}-\mathbf{x}_1}^{-\alpha_{ \sf s}} \nonumber\\ &~~~~~~+\sum_{\mathbf{b}_{m}\in \Phi_{\sf b} \cap \mathcal{A}_{\sf b}|s}G^{\sf b|s} P_{\sf b} H_{m}^{\sf s} \norm{\mathbf{b}_{m}-\mathbf{x}_1}^{-\alpha_{\sf s}} + \sigma^2\bigg)\Bigg)\nonumber\\&~~~~~~~~~~~~~~~~~~~~~~~~~~~~~~~\Bigg|\norm{\mathbf{u}_1^{ \sf s}-\mathbf{x}_1}=r, \Phi_{\sf u_{ \sf s}}\left(\mathcal{A}_{\sf u_{ \sf s}|s}\right)\ge 1\Bigg] \nonumber\\ &= \mathbb{E}_{I_{\sf up}^{\sf dn}(r)}\Bigg[\prod_{\mathbf{u}_{\ell}^{ \sf s} \in \Phi_{\sf u_s}\cap \mathcal{A}^c_{\sf u_{ \sf s}|s}(r)} \exp\left({-sG_{\ell}^{\sf u_s|s} P_{\sf u_s}  H^{ \sf s}_{\ell} \norm{\mathbf{u}_{\ell}^{ \sf s}-\mathbf{x}_1}^{-\alpha_{ \sf s}}}\right) \nonumber\\ &~~~~~~ \prod_{\mathbf{b}_{m}\in \Phi_{\sf b} \cap \mathcal{A}_{\sf b|s}} \exp\left({-sG^{\sf b|s} P_{\sf b} H^{\sf s}_{m} \norm{\mathbf{b}_{m}-\mathbf{x}_1}^{-\alpha_{\sf s}}-s  \sigma^2}\right) \nonumber\\&~~~~~~~~~~~~~~~~~~~~~~~~~~~~~~~ \Bigg|\norm{\mathbf{u}_1^{\sf s}-\mathbf{x}_1}=r, \Phi_{\sf u_{ \sf s}}\left(\mathcal{A}_{\sf u_{ \sf s}|s}\right)\ge 1\Bigg] 
        \nonumber\\ &\overset{(a)}{=} \exp\bigg({-\lambda_{\sf u_{ \sf s}} \int_{v\in \mathcal{A}^c_{\sf u_{ \sf s}|s}(r)}\left(1-\mathbb{E}_{H_{\ell}^{ \sf s}}\left[e^{-sG_{\ell}^{\sf u_s|s} P_{\sf u_s} H^{ \sf s}_{\ell}v^{-\alpha_{ \sf s}}}\right]\right)\mathrm{d}v} \nonumber\\ &~~~~{-\lambda_{\sf b}\int_{w \in \mathcal{A}_{\sf b|s}}\left(1-\mathbb{E}_{H_m^{\sf s}}\left[e^{-sG^{\sf b|s} P_{\sf b} H^{\sf s}_{m}w^{-\alpha_{\sf s}}}\right]\right)\mathrm{d}w}-s \sigma^2 \bigg) \nonumber\\&\overset{(b)}{=} \exp\bigg({-2\pi \tilde{\lambda}_{\sf u_{ \sf s}|s } \int_{r}^{R_{\sf max}^{\sf u_{ \sf s} |s}}\left(1-\mathbb{E}_{H_{\ell}^{ \sf s}}\left[e^{-sG_{\ell}^{\sf u_s|s} P_{\sf u_s} H^{ \sf s}_{\ell}v^{-\alpha_{ \sf s}}}\right]\right)v\mathrm{d}v} \nonumber\\ &- s\sigma^2 {-2\pi \tilde{\lambda}_{\sf b|s}\! \int_{R_{\sf min}^{\sf b|s}}^{R_{\psi'}^{\sf b|s}}\!\!\left(1\!-\!\mathbb{E}_{H_m^{\sf s}}\left[e^{-sG^{\sf b_L|s} P_{\sf b} H^{\sf s}_{m}w^{-\alpha_{\sf s}}}\right]\right)\!w\mathrm{d}w}    \nonumber\\ &{-2\pi \tilde{\lambda}_{\sf b|s}\! \int_{R_{\psi'}^{\sf b|s}}^{R_{\sf max}^{\sf b|s}}\!\!\left(1\!-\!\mathbb{E}_{H_m^{\sf s}}\left[e^{-sG^{\sf b_H|s} P_{\sf b} H^{\sf s}_{m}u^{-\alpha_{\sf s}}}\right]\right)\!u\mathrm{d}u} \bigg), \label{pf:laplace_updn}
    \end{align}
    where (a) follows from the probability generating functional (PGFL) of the PPP \cite{Andrews2011} and (b) results from converting the surface area into polar coordinates.
    
     By assuming $m_{\sf s}$ is integer, the PDF of $H_{\ell}^{\sf s}$ is obtained as in \cite{An2016}
    \begin{align}
        f_{H_{\ell}^{\sf s}}(x) = \sum_{z=0}^{m_{\sf s}-1}\zeta(z)x^z\exp(-(\beta-c)x).
    \end{align}
    Then, the expectations in \eqref{pf:laplace_updn} are given by
    \begin{align}
        &\mathbb{E}_{H_{\ell}^{ \sf s}}\left[e^{-sG_{\ell}^{\sf u_s|s} P_{\sf u_s} H^{ \sf s}_{\ell}v^{-\alpha_{ \sf s}}}\right] \nonumber\\ &= \int_{0}^{\infty}e^{-sG_{\ell}^{\sf u_s|s} P_{\sf u_s} v^{-\alpha_{ \sf s}}x} f_{H_\ell^{\sf s}}(x) \mathrm{d}x \nonumber\\ &=\sum_{z=0}^{m_{\sf s}-1}\zeta(z) \int_{0}^{\infty}e^{-sG_{\ell}^{\sf u_s|s} P_{\sf u_s} v^{-\alpha_{\sf s}}x} x^z e^{-(\beta-c)x}\mathrm{d}x \nonumber\\ &=\sum_{z=0}^{m_{\sf s}-1}\zeta(z) \frac{\Gamma(z+1)}{(\beta-c+sG_{\ell}^{\sf u_s|s} P_{\sf u_s} v^{-\alpha_{ \sf s}})^{z+1}},   \label{eq:laplace_expactation1}
    \end{align}
    and
    \begin{align}
        &\mathbb{E}_{H_m^{\sf s}}\left[e^{-sG^{\sf b|s} P_{\sf b} H^{\sf s}_{m}w^{-\alpha_{\sf s}}}\right] \nonumber\\ &=\sum_{i=0}^{m_{\sf s}-1}\zeta(i) \frac{\Gamma(i+1)}{(\beta-c+s G^{\sf b|s} P_{\sf b} w^{-\alpha_{\sf s}})^{i+1}}.  \label{eq:laplace_expactation2}
    \end{align}
    We finally obtain the expression for the conditional Laplace transform of the aggregated interference power plus noise power as in \eqref{eq:lem:Laplace_updn}.
\label{append:lem:Laplace_updn}
\end{proof}

\section{Proof of Theorem \ref{theorem:SE_up}}
\begin{proof}
The UL SINR coverage probability is expressed as
    \begin{align}
        &\mathbb{P}\left[{\sf SINR}_{\sf up}^{o} \ge \gamma \right] \nonumber \\& = \mathbb{P}[\Phi_{\sf u_{\sf s}}\left(\mathcal{A}_{\sf u_{\sf s}|s}\right)\ge 1] \mathbb{P}\left[{\sf SINR}_{\sf up}^{o} \ge \gamma |\Phi_{\sf u_{\sf s}}\left(\mathcal{A}_{\sf u_{\sf s}|s}\right)\ge 1\right].
    \end{align}

    The SINR coverage probability conditioned on $\Phi_{\sf u_{\sf s}}\left(\mathcal{A}_{\sf u_{\sf s}|s}\right)\ge 1$ is given by
\begin{align}
    & \mathbb{P}\left[{\sf SINR}_{\sf up}^{o} \ge \gamma |\Phi_{\sf u_{\sf s}}\left(\mathcal{A}_{\sf u_{\sf s}|s}\right)\ge 1\right]\nonumber \\ &= \mathbb{E}_{r}\Bigg[\mathbb{P}\bigg[\frac{G_{1}^{\sf u_s|s} P_{\sf u_s} H_1^{\sf s} r^{-\alpha_{\sf s}}}{I_{\sf up}^{o}(r)
    } \ge \gamma\nonumber\\&~~~~~~~~~~\bigg|\norm{\mathbf{u}_1^{\sf s}-\mathbf{x}_1}=r, \Phi_{\sf u_s}(\mathcal{A}_{\sf u_s|s})\ge 1\bigg]\Bigg|\Phi_{\sf u_s}(\mathcal{A}_{\sf u_s|s})\ge 1\Bigg] \nonumber \\  &= \mathbb{E}_{r}\bigg[\mathbb{P}\bigg[H_1^{\sf s} \ge \gamma r^{\alpha_{\sf s}}I_{\sf up}^{o}(r)\left(G_{1}^{\sf u_s|s} P_{\sf u_s}\right)^{-1}\nonumber\\&~~~~~~~~~~\Big|\norm{\mathbf{u}_1^{\sf s}-\mathbf{x}_1}=r, \Phi_{\sf u_s}(\mathcal{A}_{\sf u_s|s})\ge 1\bigg]\bigg|\Phi_{\sf u_s}(\mathcal{A}_{\sf u_s|s})\ge 1\bigg]\nonumber\\ &=\mathbb{E}_{r}\vast[\mathbb{E}_{I_{\sf up}^{o}(r)}\!\vast[1\!-\!\sum_{z=0}^{m_{\sf s}-1}\frac{\zeta(z)\Gamma(z)}{(\beta\!-\!c)^{z+1}}\!\vast(1 -\sum_{v=0}^{z} \frac{((\beta-c)\gamma r^{\alpha_{\sf s}})^v}{\left(G_{1}^{\sf u_s|s} P_{\sf u_s}\right)^{v}v!} \nonumber\\&~~ \cdot(I_{\sf up}^{o}(r))^v e^{-(\beta-c)\gamma r^{\alpha_{\sf s}} I_{\sf up}^{o}(r)\left(G_{1}^{\sf u_s|s} P_{\sf u_s}\right)^{-1}}\vast)\vast]\!\vast|\Phi_{\sf u_s}(\mathcal{A}_{\sf u_s|s})\ge 1\!\vast] \nonumber \\ 
    &\overset{(a)}{=} \mathbb{E}_{r}\vast[1-\sum_{z=0}^{m_{\sf s}-1}\!\frac{\zeta(z)\Gamma(z)}{(\beta\!-\!c)^{z+1}}\!\vast(1 -\sum_{v=0}^{z} \frac{((\beta-c)\gamma r^{\alpha_{\sf s}})^v}{\left(G_{1}^{\sf u_s|s} P_{\sf u_s}\right)^{v}v!} (-1)^v\nonumber\\&\cdot \frac{\mathrm{d}^v \mathcal{L}_{I_{\sf up}^{o}(r)|\Phi_{\sf u_s}(\mathcal{A}_{\sf u_s|s})\ge 1}(s)}{\mathrm{d}s^v} \bigg|_{ s= \frac{(\beta-c)\gamma r^{\alpha_{\sf s}}}{G_{1}^{\sf u_s|s} P_{\sf u_s}}  } \vast)\vast| \Phi_{\sf u_s}(\mathcal{A}_{\sf u_s|s})\ge 1\vast] \nonumber \\ 
    &\overset{(b)}{=} \int_{R_{\sf min}^{\sf u_s|s}}^{R_{\sf max}^{\sf u_s|s}} \vast[1-\sum_{z=0}^{m_{\sf s}-1}\frac{\zeta(z)\Gamma(z)}{(\beta\!-\!c)^{z+1}}\vast(1 -\sum_{v=0}^{z} \frac{((\beta-c)\gamma r^{\alpha_{\sf s}})^v}{\left(G_{1}^{\sf u_s|s} P_{\sf u_s}\right)^{v}v!}\nonumber\\&~~~~~~\cdot(-1)^v\frac{\mathrm{d}^v \mathcal{L}_{I_{\sf up}^{o}(r)|\Phi_{\sf u_s}(\mathcal{A}_{\sf u_s|s})\ge 1}(s)}{\mathrm{d}s^v} \bigg|_{ s= \frac{(\beta-c)\gamma r^{\alpha_{\sf s}}}{G_{1}^{\sf u_s|s} P_{\sf u_s}}  } \vast)\vast] \nonumber\\ &~~~~~~~~~~~~~~~~~~~\cdot f_{\norm{\mathbf{u}_1^{\sf s}-\mathbf{x}_1}|\Phi_{\sf u_{\sf s}}(\mathcal{A}_{\sf u_{\sf s}|s})\ge 1}(r) \mathrm{d}r,
\end{align}
where (a) follows from applying the derivative property of the Laplace transform, i.e., $\mathbb{E}\left[X^v e^{-sX}\right]=(-1)^v \frac{\mathrm{d}\mathcal{L}_X(s)}{\mathrm{d}s^v}$, and (b) comes from the expectation over the PDF obtained in Lemma \ref{lem:nearest_distance}. Then, by marginalizing this coverage probability over $\gamma$, we obtain the ergodic spectral efficiency expression in \eqref{eq:theorem1}.

\label{append:theorem:SE_up}
\end{proof}

\section{Proof of Corollary \ref{cor:LB_UL}}
\begin{proof}
The ergodic spectral efficiency of the UL satellite networks with spectrum sharing with terrestrial networks is lower bounded as

    \begin{align}      
        &R_{\sf up}^{o} =\mathbb{E}\left[\log_2\left(1+\frac{G_{1}^{\sf u_s|s} P_{\sf u_s} H_1^{\sf s} \norm{\mathbf{u}_1^{\sf s}-\mathbf{x}_1}^{-2}}{I_{\sf s}^{\sf up} +I_{\sf t,up}^{o} + \sigma^2}\right)\right] \nonumber\\&
        \overset{(a)}{\ge} \log_2\left(1+\frac{\exp\left(\mathbb{E}\left[\ln \left(G_{1}^{\sf u_s|s} P_{\sf u_s} H_1^{\sf s} \norm{\mathbf{u}_1^{\sf s}-\mathbf{x}_1}^{-2}\right)\right]\right)}{\mathbb{E}\left[I_{\sf s}^{\sf up} +I_{\sf t,up}^{o} + \sigma^2\right]}\right) \nonumber\\ &= \log_2\left(1+\frac{\exp\left(\mathbb{E}\left[\ln \left(G_{1}^{\sf u_s|s} P_{\sf u_s} H_1^{\sf s} \norm{\mathbf{u}_1^{\sf s}-\mathbf{x}_1}^{-2}\right)\right]\right)}{\mathbb{E}\left[I_{\sf s}^{\sf up}\right] +\mathbb{E} \left[I_{\sf t,up}^{o}\right] + \mathbb{E}\left[\sigma^2\right]}\right), \label{eq:LBforup}
    \end{align}
    where (a) follows from the Lemma 2 in \cite{Lee2016-2}.
 
Therefore, the lower bound of the ergodic spectral efficiency of UL satellite networks for UL and DL terrestrial networks differs in term $\mathbb{E}\left[I_{\sf t,up}^o\right]$. The expectation of the aggregated interference power from DL terrestrial networks to the UL satellite networks is computed by
\begin{align}
    \mathbb{E}\left[I_{\sf t,up}^{\sf dn}\right] &= \mathbb{E}\left[\sum_{\mathbf{b}_{m}\in \Phi_{\sf b} \cap \mathcal{A}_{\sf b|s}}G^{\sf b|s} P_{\sf b} H_{m}^{\sf s} \norm{\mathbf{b}_{m}-\mathbf{x}_1}^{-2}\right] \nonumber\\ &\overset{(a)}{=} 2\pi \tilde{\lambda}_{\sf b|s}\int_{R_{\sf min}^{\sf b|s}}^{R_{\psi'}^{\sf b|s}}v \left(G^{\sf b_L|s} P_{\sf b}\mathbb{E}\left[H_{m}^{\sf s}\right]v^{-2}\right) \mathrm{d}v \nonumber\\ &+2\pi \tilde{\lambda}_{\sf b|s}\int_{R_{\psi'}^{\sf b|s}}^{R_{\sf max}^{\sf b|s}}w \left(G^{\sf b_H|s} P_{\sf b}\mathbb{E}\left[H_{m}^{\sf s}\right]w^{-2}\right) \mathrm{d}w \nonumber\\ 
    &=2\pi \tilde{\lambda}_{\sf b|s} G^{\sf b_L|s}P_{\sf b}\left(\ln\left(R_{\psi'}^{\sf b|s}\right)-\ln\left(R_{\sf min}^{\sf b|s}\right)\right) \nonumber\\ &+2\pi \tilde{\lambda}_{\sf b|s} G^{\sf b_H|s}P_{\sf b}\left(\ln\left(R_{\sf max}^{\sf b|s} \right)-\ln\left(R_{\psi'}^{\sf b|s}\right)\right), \label{eq:LB_updn}
\end{align}
where (a) is obtained by applying Campbell's theorem \cite{Baccelli2009}.
Similarly, the expectation of the aggregated interference power from UL terrestrial networks to the UL satellite networks is computed as follows: 
\begin{align}
     \mathbb{E}\left[I_{\sf t,up}^{\sf up}\right] &= \mathbb{E}\left[\sum_{\mathbf{u}_{k}^{\sf t} \in \Phi_{\sf u_{\sf t}}\cap \mathcal{A}_{\sf u_{\sf t}|s}}G^{\sf u_t|s} P_{\sf u_t}  H^{\sf s}_{k} \norm{\mathbf{u}_{k}^{\sf t}-\mathbf{x}_1}^{-2}\right] \nonumber\\ &=2\pi \tilde{\lambda}_{\sf u_t|s} G^{\sf u_t|s}P_{\sf u_t}\left(\ln\left(R_{\sf max }^{\sf u_t|s}\right)-\ln\left(R_{\sf min}^{\sf u_t|s}\right)\right).  \label{eq:LB_upup}
\end{align}

By comparing \eqref{eq:LB_updn} and \eqref{eq:LB_upup}, we obtain the condition that the DL terrestrial networks introduce more interference to the UL satellite networks than the UL terrestrial networks as follows:
\begin{align}
    \frac{\lambda_{\sf u_t}}{\lambda_{\sf b}}\le\frac{P_{\sf b}R_{\sf b}}{P_{\sf u_t}R_{\sf u_t}}\frac{G^{\sf b_L|s}\ln\left(\frac{R_{\psi'}^{\sf b|s}}{R_{\sf min}^{\sf b|s}}\right)+G^{\sf b_H|s}\ln\left(\frac{R_{\sf max}^{\sf b|s}}{R_{\psi'}^{\sf b|s}}\right)}{G^{\sf u_t|s}\ln\left(\frac{R_{\sf max}^{\sf u_t|s}}{R_{\sf min}^{\sf u_t|s}}\right)}.
\end{align}
   \label{append:cor:LB_UL}
\end{proof}

\section{Proof of Lemma \ref{lem:Laplace_dndn}}
\begin{proof}
Let $\mathcal{A}^c_{\sf u_{ \sf s}}(r)=\mathcal{A}_{\sf u_{ \sf s}}/\mathcal{A}_{\sf u_{ \sf s}}(r)$. Then, the conditional Laplace transform for the aggregated interference power plus noise power is computed as
    \begin{align}
        &\mathcal{L}_{I_{\sf dn}^{\sf dn}(r)|\Phi_{\sf s}\left(\mathcal{A}_{\sf s|u_{ \sf s}}\right)\ge 1}(s) \nonumber\\&= \mathbb{E}_{I_{\sf dn}^{\sf dn}(r)}\!\left[\exp(-sI_{\sf dn}^{\sf dn}(r))\big|\Phi_{\sf s}\!\left(\mathcal{A}_{\sf s|u_s}\right)\!\ge\! 1\right] 
        % \nonumber\\&=\mathbb{E}_{I_{\sf dn}^{\sf dn}(r)}\!\Bigg[\!\exp\!\Bigg(\!\!\!-\!s\bigg( \sum_{\mathbf{x}_n \in \Phi_{\sf s}\cap \mathcal{A}^c_{\sf s|u_s}(r)}\!\!G_{n}^{\sf s|u_s} P_{\sf s}  H^{ \sf s}_{n} \norm{\mathbf{x}_n-\mathbf{u}_1^{ \sf s}}^{-\alpha_{ \sf s}} \nonumber\\ &~~~~~~+\sum_{\mathbf{b}_{m}\in \Phi_{\sf b} \cap \mathcal{A}_{\sf b|u_s}}G^{\sf b| u_s} P_{\sf b} H_{m}^{\sf t} \norm{\mathbf{b}_{m}-\mathbf{u}_1^{ \sf s}}^{-\alpha_{\sf t}} + \sigma^2\bigg)\Bigg)\nonumber\\&~~~~~~~~~~~~~~~~~~~~~~~~~~~~~~~\Bigg|\norm{\mathbf{x}_1-\mathbf{u}_1^{\sf s}}=r, \Phi_{\sf s}\left(\mathcal{A}_{\sf s|u_s}\right)\ge 1\Bigg] \nonumber\\ &= \mathbb{E}_{I_{\sf dn}^{\sf dn}(r)}\Bigg[\prod_{\mathbf{x}_{n} \in \Phi_{\sf s}\cap \mathcal{A}^c_{\sf s|u_s}(r)} \exp\left({-sG_{n}^{\sf s|u_s} P_{\sf s}  H^{ \sf s}_{n} \norm{\mathbf{x}_{n}-\mathbf{u}_1^{\sf s}}^{-\alpha_{ \sf s}}}\right) \nonumber\\ &~~~~~~ \prod_{\mathbf{b}_{m}\in \Phi_{\sf b} \cap \mathcal{A}_{\sf b|u_s}} \exp\left({-sG^{\sf b|u_s} P_{\sf b} H^{\sf t}_{m} \norm{\mathbf{b}_{m}-\mathbf{u}_1^{\sf s}}^{-\alpha_{\sf t}} -s\sigma^2}\right) \nonumber\\&~~~~~~~~~~~~~~~~~~~~~~~~~~~~~~~ \Bigg|\norm{\mathbf{x}_1-\mathbf{u}_1^{\sf s}}=r, \Phi_{\sf s}\left(\mathcal{A}_{\sf s|u_s}\right)\ge 1\Bigg] 
        \nonumber\\ &\overset{(a)}{=} \exp\bigg({-\lambda_{\sf s} \int_{v\in \mathcal{A}^c_{\sf s|u_s}(r)}\left(1-\mathbb{E}_{H_{n}^{ \sf s}}\left[e^{-sG_{n}^{\sf s|u_s} P_{\sf s} H^{ \sf s}_{n}v^{-\alpha_{ \sf s}}}\right]\right)\mathrm{d}v} \nonumber\\ &{-\lambda_{\sf b}\int_{w \in \mathcal{A}_{\sf b|u_s}}\left(1-\mathbb{E}_{H_m^{\sf t}}\left[e^{-sG^{\sf b |u_s} P_{\sf b} H^{\sf t}_{m}w^{-\alpha_{\sf t}}}\right]\right)\mathrm{d}w}-s \sigma^2 \bigg) \nonumber\\&\overset{(b)}{=} \exp\bigg({-2\pi \tilde{\lambda}_{\sf s|u_s} \int_{r}^{R_{\sf max}^{\sf s|u_{ \sf s} }}\left(1\!-\!\mathbb{E}_{H_{n}^{ \sf s}}\left[e^{-sG_{n}^{\sf s|u_s} P_{\sf s} H^{ \sf s}_{n}v^{-\alpha_{ \sf s}}}\right]\right)v\mathrm{d}v} \nonumber\\ &- s\sigma^2 \!{-2\pi \tilde{\lambda}_{\sf b|u_s}\! \int_{R_{\sf min}^{\sf b|u_s}}^{R_{\sf max}^{\sf b|u_s}}\!\!\left(1\!-\!\mathbb{E}_{H_m^{\sf t}}\left[e^{-sG^{\sf b_H|u_s} P_{\sf b} H^{\sf t}_{m}w^{-\alpha_{\sf t}}}\right]\right)\!w\mathrm{d}w}  \bigg)\nonumber\\&=\exp\Bigg(- s\sigma^2 \nonumber\\ & {-2\pi \tilde{\lambda}_{\sf s|u_s} \int_{r}^{R_{\sf max}^{\sf s|u_{ \sf s} }}\left[1-\sum_{z=0}^{m_{\sf s}-1}\! \frac{\zeta(z)\Gamma(z+1)}{(\beta-c+sG_{\ell}^{\sf s|u_s} P_{\sf s} v^{-\alpha_{ \sf s}})^{z+1}}\right]v\mathrm{d}v} \nonumber\\ & {-2\pi \tilde{\lambda}_{\sf b|u_s}\! \int_{R_{\sf min}^{\sf b|u_s}}^{R_{\sf max}^{\sf b|u_s}}\!\!\left[1\!-\!{\left(1+\frac{sG^{\sf b_H|u_s}P_{\sf b}w^{-\alpha_{\sf t}}}{m_{\sf t}}\right)^{-m_{\sf t}}}\right]\!w\mathrm{d}w}  \Bigg), \label{pf:laplace_dndn}
    \end{align}
    where (a) is derived from the PGFL of the PPP \cite{Andrews2011} and (b) is due to $\sqrt{H_m^{\sf t}}$ following the Nakagami distribution.

    \label{append:lem:Laplace_dndn}
\end{proof}

\section{Proof of Theorem \ref{theorem:SE_dn}}
\begin{proof}
The DL SINR coverage probability is expressed as
    \begin{align}
        &\mathbb{P}\left[{\sf SINR}_{\sf dn}^{o} \ge \gamma \right] \nonumber \\& = \mathbb{P}[\Phi_{\sf s}\left(\mathcal{A}_{\sf s|u_s}\right)\ge 1] \mathbb{P}\left[{\sf SINR}_{\sf dn}^{o} \ge \gamma |\Phi_{\sf s}\left(\mathcal{A}_{\sf s|u_s}\right)\ge 1\right].
    \end{align}

    The SINR coverage probability conditioned on $\Phi_{\sf s}\left(\mathcal{A}_{\sf s|u_s}\right)\ge 1$ is given by
\begin{align}
    & \mathbb{P}\left[{\sf SINR}_{\sf dn}^{o} \ge \gamma |\Phi_{\sf s}\left(\mathcal{A}_{\sf s|u_s}\right)\ge 1\right]\nonumber \\ &= \mathbb{E}_{r}\Bigg[\mathbb{P}\bigg[\frac{G_{1}^{\sf s|u_s} P_{\sf s} H_1^{\sf s} r^{-\alpha_{\sf s}}}{I_{\sf dn}^{o}(r)
    } \ge \gamma\nonumber\\&~~~~~~~~~~\bigg|\norm{\mathbf{x}_1-\mathbf{u}_1^{\sf s}}=r, \Phi_{\sf s}(\mathcal{A}_{\sf s|u_s})\ge 1\bigg]\Bigg|\Phi_{\sf s}(\mathcal{A}_{\sf s|u_s})\ge 1\Bigg] \nonumber \\  &= \mathbb{E}_{r}\bigg[\mathbb{P}\bigg[H_1^{\sf s} \ge \gamma r^{\alpha_{\sf s}}I_{\sf dn}^{o}(r)\left(G_{1}^{\sf s|u_s} P_{\sf s}\right)^{-1}\nonumber\\&~~~~~~~~~~\Big|\norm{\mathbf{x}_1-\mathbf{u}_1^{\sf s}}=r, \Phi_{\sf s}(\mathcal{A}_{\sf s|u_s})\ge 1\bigg]\bigg|\Phi_{\sf s}(\mathcal{A}_{\sf s|u_s})\ge 1\bigg]\nonumber\\ &=\mathbb{E}_{r}\vast[\mathbb{E}_{I_{\sf dn}^{o}(r)}\!\vast[1\!-\!\sum_{z=0}^{m_{\sf s}-1}\frac{\zeta(z)\Gamma(z)}{(\beta\!-\!c)^{z+1}}\!\vast(1 -\sum_{v=0}^{z} \frac{((\beta-c)\gamma r^{\alpha_{\sf s}})^v}{\left(G_{1}^{\sf s|u_s} P_{\sf s}\right)^{v}v!} \nonumber\\&~~ \cdot(I_{\sf dn}^{o}(r))^v e^{-(\beta-c)\gamma r^{\alpha_{\sf s}} I_{\sf dn}^{o}(r)\left(G_{1}^{\sf s|u_s} P_{\sf s}\right)^{-1}}\vast)\vast]\!\vast|\Phi_{\sf s}(\mathcal{A}_{\sf s|u_s})\ge 1\!\vast] \nonumber \\ 
    &= \mathbb{E}_{r}\vast[1-\sum_{z=0}^{m_{\sf s}-1}\!\frac{\zeta(z)\Gamma(z)}{(\beta\!-\!c)^{z+1}}\!\vast(1 -\sum_{v=0}^{z} \frac{((\beta-c)\gamma r^{\alpha_{\sf s}})^v}{\left(G_{1}^{\sf s|u_s} P_{\sf s}\right)^{v}v!} (-1)^v \nonumber\\&\cdot \frac{\mathrm{d}^v \mathcal{L}_{I_{\sf dn}^{o}(r)|\Phi_{\sf s}(\mathcal{A}_{\sf s|u_s})\ge 1}(s)}{\mathrm{d}s^v} \bigg|_{ s= \frac{(\beta-c)\gamma r^{\alpha_{\sf s}}}{G_{1}^{\sf s|u_s} P_{\sf s}}  } \vast)\vast| \Phi_{\sf s}(\mathcal{A}_{\sf s|u_s})\ge 1\vast] \nonumber \\ 
    &\overset{(a)}{=} \int_{R_{\sf min}^{\sf s|u_s}}^{R_{\sf max}^{\sf s|u_s}} \vast[1-\sum_{z=0}^{m_{\sf s}-1}\frac{\zeta(z)\Gamma(z)}{(\beta\!-\!c)^{z+1}}\vast(1 -\sum_{v=0}^{z} \frac{((\beta-c)\gamma r^{\alpha_{\sf s}})^v}{\left(G_{1}^{\sf s|u_s} P_{\sf s}\right)^{v}v!}\nonumber\\&~~~~~~\cdot(-1)^v\frac{\mathrm{d}^v \mathcal{L}_{I_{\sf dn}^{o}(r)|\Phi_{\sf s}(\mathcal{A}_{\sf s|u_s})\ge 1}(s)}{\mathrm{d}s^v} \bigg|_{ s= \frac{(\beta-c)\gamma r^{\alpha_{\sf s}}}{G_{1}^{\sf s|u_s} P_{\sf s}}  } \vast)\vast] \nonumber\\ &~~~~~~~~~~~~~~~~~~~\cdot f_{\norm{\mathbf{x}_1-\mathbf{u}_1^{\sf s}}|\Phi_{\sf s}(\mathcal{A}_{\sf s|u_s})\ge 1}(r) \mathrm{d}r, 
\end{align}
where (a) comes from the expectation over the PDF obtained in Lemma \ref{lem:nearest_sat_distance}. 
By integrating this coverage probability over $\gamma$, we obtain the expression for ergodic spectral efficiency as given in \eqref{eq:theorem:SE_dn}.
\label{append:theorem:SE_dn}
\end{proof}

\section{Proof of Corollary \ref{cor:LB_DL}}
\begin{proof}
The ergodic spectral efficiency of the DL satellite networks with spectrum sharing with terrestrial networks is lower bounded as
    \begin{align}      
        &R_{\sf dn}^{o} =\mathbb{E}\left[\log_2\left(1+\frac{G_{1}^{\sf s|u_s} P_{\sf s} H_1^{\sf s} \norm{\mathbf{x}_1-\mathbf{u}_1^{\sf s}}^{-2}}{I_{\sf s}^{\sf dn} +I_{\sf t,dn}^{o} + \sigma^2}\right)\right] \nonumber\\&
       \overset{(a)}{\le}\log_2\left(1+\frac{\exp\left(\mathbb{E}\left[\ln \left(G_{1}^{\sf s|u_s} P_{\sf s} H_1^{\sf s} \norm{\mathbf{x}_1-\mathbf{u}_1^{\sf s}}^{-2}\right)\right]\right)}{\mathbb{E}\left[I_{\sf s}^{\sf dn}\right] +\mathbb{E} \left[I_{\sf t,dn}^{o}\right] + \mathbb{E}\left[\sigma^2\right]}\right), \label{eq:LBfordn}
    \end{align}
    where (a) follows from the Lemma 2 in \cite{Lee2016-2}.
 
Therefore, the lower bound of the ergodic spectral efficiency of DL satellite networks for UL and DL terrestrial networks differs in term $\mathbb{E}\left[I_{\sf t,dn}^o\right]$. The expectation of the aggregated interference power from DL terrestrial networks to the DL satellite networks is computed by
\begin{align}
    \mathbb{E}\left[I_{\sf t,dn}^{\sf dn}\right] &= \mathbb{E}\left[\sum_{\mathbf{b}_{m}\in \Phi_{\sf b} \cap \mathcal{A}_{\sf b|u_s}}G^{\sf b|u_s} P_{\sf b} H_{m}^{\sf t} \norm{\mathbf{b}_{m}-\mathbf{u}_1^{\sf s}}^{-4}\right] \nonumber\\ &\overset{(a)}{=} 2\pi \tilde{\lambda}_{\sf b|u_s}\int_{R_{\sf min}^{\sf b|u_s}}^{R_{\sf max}^{\sf b|u_s}}v \left(G^{\sf b_H|u_s} P_{\sf b}\mathbb{E}\left[H_{m}^{\sf t}\right]v^{-4}\right) \mathrm{d}v \nonumber\\
    &=\pi \tilde{\lambda}_{\sf b|u_s} G^{\sf b_H|u_s}P_{\sf b}\left(\frac{1}{\left(R_{\sf min}^{\sf b|u_s}\right)^2}-\frac{1}{\left(R_{\sf max}^{\sf b|u_s}\right)^2}\right), \label{eq:LB_dndn}
\end{align}
where (a) is obtained by applying Campbell's theorem \cite{Baccelli2009}.
Similarly, the expectation of the aggregated interference power from UL terrestrial networks to the DL satellite networks is computed as follows: 
\begin{align}
     \mathbb{E}\left[I_{\sf t,dn}^{\sf up}\right] &= \mathbb{E}\left[\sum_{\mathbf{u}_{k}^{\sf t} \in \Phi_{\sf u_{\sf t}}\cap \mathcal{A}_{\sf u_{\sf t}|u_s}}G^{\sf u_t|u_s} P_{\sf u_t}  H^{\sf t}_{k} \norm{\mathbf{u}_{k}^{\sf t}-\mathbf{u}_1^{\sf s}}^{-4} \right] \nonumber\\ &=\pi {\lambda}_{\sf u_t|u_s} G^{\sf u_t|u_s}P_{\sf u_t}\left(\frac{1}{\left(R_{\sf min}^{\sf u_t|u_s}\right)^2}-\frac{1}{\left(R_{\sf max}^{\sf u_t|u_s}\right)^2}\right).  \label{eq:LB_dnup}
\end{align}

By comparing \eqref{eq:LB_dndn} and \eqref{eq:LB_dnup}, we obtain the condition that the DL terrestrial networks introduce more interference to the DL satellite networks than the UL terrestrial networks as follows:
\begin{align}
    \frac{\lambda_{\sf u_t}}{\lambda_{\sf b}}\le\frac{P_{\sf b}R_{\sf b}G^{\sf b_H|s}}{P_{\sf u_t}R_{\sf u_s}G^{\sf u_t|u_s}}\frac{\frac{1}{\left(R_{\sf min}^{\sf b|u_s}\right)^2}-\frac{1}{\left(R_{\sf max}^{\sf b|u_s}\right)^2}}{\frac{1}{\left(R_{\sf min}^{\sf u_t|u_s}\right)^2}-\frac{1}{\left(R_{\sf max}^{\sf u_t|u_s}\right)^2}}.
\end{align}
   \label{append:cor:LB_DL}
\end{proof}

\end{appendices}

\bibliographystyle{IEEEtran}
\bibliography{IEEEabrv,Reference}

\end{document}